\documentclass[pdftex,epsf]{iopart}
\usepackage{iopams,amsfonts,rotating,graphicx,fancybox,fancyhdr
,dsfont}
\usepackage{mathrsfs,amssymb,amsthm,amscd,amsfonts,bm,color,setstack}

\usepackage[hyperfootnotes=true,
        pdffitwindow=true,
        plainpages=false,
        pdfpagelabels=true,
        pdfpagemode=UseOutlines,
        pdfpagelayout=SinglePage,
        hyperindex,]{hyperref}

\newcommand{\F}{{\rm F}}
\newcommand{\B}{{\rm B}}
\newcommand{\U}{{\rm U}}
\newcommand{\diag}{{\,\rm diag}}
\newcommand{\id}{\mathbf{1}}
\newcommand{\Str}{{\,\rm Str\,}}
\newcommand{\Sdet}{{\,\rm Sdet\,}}
\newcommand{\Herm}{{\,\rm Herm}}
\newtheorem{conjecture}{Conjecture}

\begin{document}
\title[Local Tail Statistics]{Local Tail Statistics of Heavy-Tailed Random Matrix Ensembles with Unitary Invariance}

\author{ M.\ Kieburg and A.\ Monteleone}

\address{
School of Mathematics and Statistics, University of Melbourne, 813 Swanston Street, Parkville, Melbourne VIC 3010, Australia
}
\ead{m.kieburg@unimelb.edu.au, monteleonea@student.unimelb.edu.au}
\begin{abstract}
We study heavy-tailed Hermitian random matrices that are unitarily invariant. The invariance implies that the eigenvalue and eigenvector statistics are decoupled. The motivating question has been whether a  freely  stable  random matrix has stable  eigenvalue  statistics  for the largest eigenvalues in the tail. We investigate this question through the use of both numerical and analytical means, the latter of which makes use of the supersymmetry method. A surprising behaviour is uncovered in that a freely stable random matrix does not necessarily yield stable statistics and if it does then it might exhibit  Poisson  or  Poisson-like statistics. The Poisson statistics have been already observed for heavy-tailed Wigner matrices.  We conclude with two conjectures on this peculiar behaviour.

\end{abstract}

\maketitle


\section{Introduction}\label{sec:intro}

Gaussian random matrices are a well-studied topic due to the enormous number of applications in engineering, mathematics and physics. This large variety in scope has led to a multitude of modifications of random matrix models veering away from the traditional Gaussian ensembles into models with greater generality, e.g. see the textbooks~\cite{Mehtabook,Peterbook,Gernotbook}. This has resulted in various studies of spectral statistics such as bulk  statistics, hard- and soft-edge statistics, as well as the statistics of multicritical points where for instance spectral supports merge (also known as cuts) or an outlier (a separate eigenvalue not belonging to the bulk of the spectrum) is absorbed into the bulk of the spectrum.

What is less well-studied is the tail statistics of eigenvalues of a random matrix that exhibits a heavy tail. Nevertheless, heavy-tailed random matrices have many important applications especially when systems are non-stationary or open. For such systems one can expect heavy-tails being more natural to occur rather than ensembles for which all moments exist. Examples admitting heavy tails being time series analysis, disordered systems, quantum field theory and more recently deep neural networks.
e.g., see~\cite{BJNPZ2001,MS2003,BGW2006,BCP2008,AFV2010,BT2012,MSG2014,Kanazawa2016,OT2017,Minsker2018,MM2018,MM2019,Heiny2019a}. From a theoretical point of view the ensembles stable under matrix addition are of particular importance since they are the fixed points of their respective domains of attraction via the multivariate central limit theorem~\cite{Rvaceva1954}, see~\cite{ZK2021} for a recent work on unitarily invariant Hermitian random matrices. The classification of these domains as well as the stable distributions is still poorly understood from the perspective of spectral statistics. We aim to unveil one part of this incomplete picture, namely the statistics of the largest eigenvalues for unitarily invariant ensembles.

There are several works~\cite{CB1994,BJNPZ2001,Heiny2019a,Soshnikov2004,BBP2007,BJNPZ2007,BAG2008,ABAP2009, Vershynin2012, BGGM2014,BGM2016,HM2017,BG2017,Male2017,GLPTJ2017}, in physics and mathematics that have studied heavy-tailed Wigner matrices in detail (meaning matrices whose entries are independently and identically distributed along a heavy-tailed univariate probability measure). For these matrices it has been shown~\cite{Soshnikov2004,BBP2007,ABAP2009,TBT2016} that the largest eigenvalues in the heavy tail converge to Poisson statistics. Moreover the eigenvectors become localised~\cite{BT2012,CB1994,TBT2016} while those in the bulk become delocalised~\cite{BG2017}. One may argue that the Poisson statistics of the eigenvalues is due to the localisation of the eigenvectors. This perspective is supported by the fact that the distribution of these largest eigenvalues are shared by the distribution of the largest matrix entries~\cite{ABAP2009}. In the present work, we will argue that this is not necessarily the case and that the Poisson statistics or at least a very diminished level repulsion can also be found for unitarily invariant random matrix ensembles. Such ensembles have been discussed in ~\cite{AFV2010,Kanazawa2016,BJNPZ2007,BJJNPZ2002,AV2008,AMAV2009,CM2009,Thomas,Tdistribution}. The unitary invariance takes the eigenvectors out of the picture as they are still Haar distributed and thus delocalised. In~\cite{Balian1968}, it was also shown that exactly such matrices maximise the Shannon entropy when only a level density is given as an input.

In our present work we consider two specific random matrix ensembles. The first one is about the singular values of a product of complex inverse Ginibre matrices, see~\cite{ARRS2016,Forrester2014,AI2015,LWZ2016} for an analytical computation of the finite $N$ (matrix dimension) statistics as well as the hard edge statistics. This ensemble is not stable for finite matrix dimension, it is known~\cite{BPB1999,APA2009} that the limiting macroscopic level density is stable under free convolution (sum of identical and independent copies of the random matrix in the large $N$ limit). Numerically we have confirmed that this asymptotic stability also holds for the local statistics at the soft-edge and some part of the bulk. Indeed in Ref.~\cite{LWZ2016}, those local spectral statistics have been proven for this kind of random matrix. However the tail statistics do not share this behaviour. In the tail the eigenvalues degenerate statistically, meaning clusters of eigenvalues show a diminished level repulsion which is vanishing completely for $N\to\infty$. This number of eigenvalues inside such a cluster is equal to the number of copies of the random matrix that has been added. Through using the supersymmetry method we have analytically confirmed our numerical observations. Namely inside the tail the sum of random matrices agrees with the direct sum of exactly the same matrices. This agreement is however not true in the bulk or at the soft-edge. Therefore there is a transition. The scaling of the eigenvalues that belong to this critical regime of the transition has been identified and exhibits a dependence on the stability exponent.

Furthermore, we will address the question of the central limit theorem for the tail statistics of the sum of these random matrices, which happens to converge to a Poisson point process for the largest eigenvalues as it has already been known for heavy-tailed Wigner matrices~\cite{Soshnikov2004,BBP2007,ABAP2009,TBT2016}. To confirm whether this picture is true more generally, we have consider a second random matrix ensemble which is a Gaussian unitary ensemble (GUE) whose variance is averaged over a stable one-sided distribution. Similar averages have also been discussed in~\cite{BGW2006,BCP2008,AFV2010,AV2008,AMAV2009}. This construction yields a random matrix ensemble that is already stable at fixed matrix dimension $N$ so that its macroscopic as well as its microscopic statistics must be stable. There is however the downside that two copies of the matrix are not free in the sense of free probability~\cite{Speicher2011}. Whether their largest eigenvalues will go exactly to the Poisson statistics will depend on whether or not the largest eigenvalues live on a scale that is bigger than the one of the bulk. This will be shown with a supersymmetry calculation. The Monte Carlo simulations we have generated suggest that the average position of the largest eigenvalues might saturate at a finite value.

The present work is built up as follows. In Sec.~\ref{sec:observe}, we describe the numerical experiment we have carried out for a sum of inverse complex Ginibre matrices in order to get a feeling what is happening. We do not only confirm that the macroscopic level density as well as the soft edge statistics are stable but show how the tail statistics change by adding several independent copies of these random matrices. To get an analytical confirmation, we compute the average of the ratio of characteristic polynomials in Sec.~\ref{sec:analytical}. Therein, we consider a more general situation of the sum of products of inverse Ginibre matrices as those are also known to be stable under free convolution~\cite{BPB1999,APA2009}. Those averages encode the whole eigenvalue statistics and are computed via the supersymmetry method~\cite{Efetovbook,Zirnbauer2006,Guhr2011}. In Sec.~\ref{sec:poisson}, we study the limit of the sum of an infinite number of matrices analytically as well as numerically. In particular we investigate the question regarding how universal are the Poisson statistics for the largest eigenvalues of heavy-tailed ensembles and what are the scales to finding them. We conclude in Sec.~\ref{sec:conclusio} by formulating two conjectures for heavy-tailed ensembles.

\section{Numerical Observations}\label{sec:observe}

We begin this section with a short numerical experiment (In fact the same one that led to the discovery of this surprising result) as it will give us some insight into what is going when we add random matrices with heavy tailed macroscopic level densities. 

We consider a sum of $L$ identically and independently distributed inverse complex Ginibre matrices $X_j\in\mathbb{C}^{N\times N}$ with the probability density
\begin{equation}\label{Inv-Gin}
P(X_j)=\frac{1}{\pi^{N^2}}\frac{\exp[-\tr (X_j^\dagger X_j)^{-1}]}{\det(X_j^\dagger X_j)^{2N}}.
\end{equation}
Certainly, the inverse $X_j^{-1}$ is a complex Ginibre matrix~\cite{Ginibre1965} and the product $(X_j^\dagger X_j)^{-1}$ describes a complex (Wishart) Laguerre matrix~\cite{Mehtabook,Wishart1928}. Thus, the full spectral statistics of a single matrix ($L=1$) is completely known, see~\cite{Mehtabook,Peterbook}.

For instance, the macroscopic level density of $(X_j^\dagger X_j)^{-1}$ is the Mar\v{c}enko-Pastur law~\cite{MP1967},
\begin{equation}\label{MP-law}
\fl\rho_{(X_j^\dagger X_j)^{-1}}(\lambda)=\lim_{N\to\infty}\left\langle\frac{1}{N}\tr\delta\left(\lambda\id_N-\frac{1}{N}(X_j^\dagger X_j)^{-1}\right)\right\rangle=\frac{\sqrt{4-\lambda}}{2\pi\sqrt{\lambda}}\Theta[\lambda(4-\lambda)]
\end{equation}
with the Heaviside step function $\Theta$.
This implies that the macroscopic level density of $X_j^\dagger X_j$ is equal to
\begin{equation}\label{invMP-law}
\fl\rho_{X_j^\dagger X_j}(\lambda)=\lim_{N\to\infty}\left\langle\frac{1}{N}\tr\delta\left(\lambda\id_N-NX_j^\dagger X_j\right)\right\rangle=\frac{\sqrt{4-\lambda^{-1}}}{2\pi\lambda^{3/2}}\Theta[\lambda(4-\lambda^{-1})].
\end{equation}
Comparing this result with the definition of the stability exponent $\alpha$, see~\cite[Appendix A]{BPB1999}, which shows in the asymptotic approximation $\rho(\lambda)\propto |\lambda|^{-1-\alpha}$ for  large $|\lambda|\gg1$, the current ensemble corresponds to $\alpha=1/2$.

\begin{figure}[t!]
\begin{center}
\includegraphics[width=0.8\textwidth]{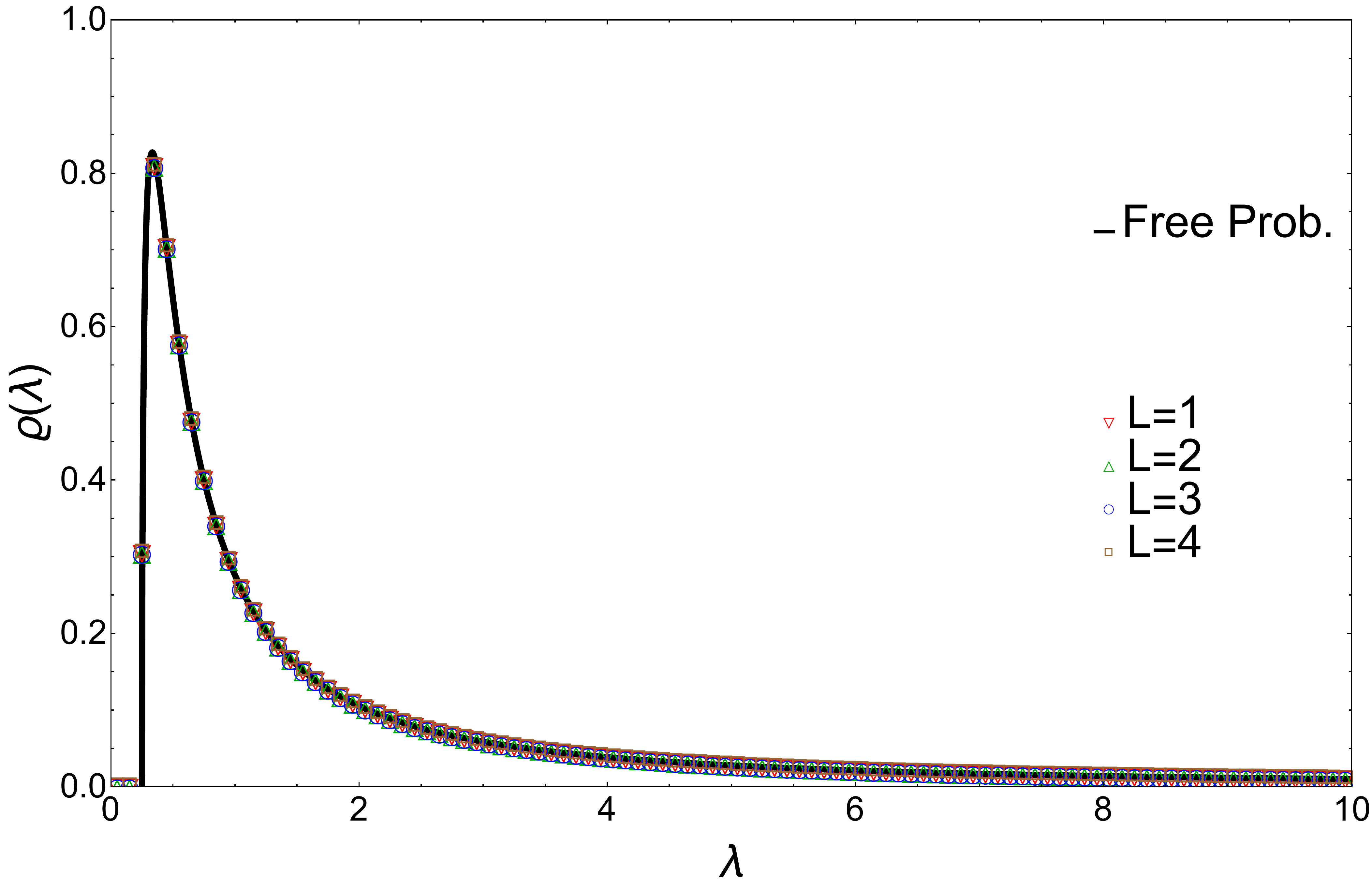}
\caption{The macroscopic level density of the random matrix sum~\eref{sum-random} has been simulated by Monte Carlo simulations (coloured symbols) for $L$ numbers of matrices added together. This is compared to the analytical result~\eref{invMP-law} (black solid curve) which should agree for any $L$ via free probability. We generated for each setting $10^6$ configurations of dimension $N=200$. The bin size is equal to $0.1$. Thus, the statistical and systematic error is below one percent.}
\label{fig:rhomacro-alp05-bet1}
\end{center}
\end{figure}

From free probability~\cite{BPB1999,APA2009}, we know that this distribution is stable under free convolution, meaning when we add two or more copies of the matrix $X_j^\dagger X_j$, i.e.,
\begin{equation}\label{sum-random}
Y_L=\frac{1}{L^{1/\alpha}}\sum_{j=1}^LX_j^\dagger X_j,
\end{equation}
 the resulting matrix $Y_L$ shares the same macroscopic level density as each single $X_j^\dagger X_j$. This can be readily checked via the $\mathcal{R}$-transform which is implicitly defined with the help of the Green function~\cite{Speicher2011}
\begin{equation}\label{Greenfunction}
G(z)=\int_{-\infty}^\infty\frac{\rho(\lambda)d\lambda}{z-\lambda},
\end{equation}
namely
\begin{equation}\label{R-transform}
\mathcal{R}[G(z)]=z-\frac{1}{G(z)}.
\end{equation}
The $R$-transform of $X_j^\dagger X_j$ is~\cite{APA2009}
\begin{equation}\label{R-trafo-invGin}
\mathcal{R}_{X_j^\dagger X_j}(y)=-e^{i\pi/2}y^{-1/2}.
\end{equation}
Due to the rule for a sum of two asymptotically free random matrices $A$ and $B$,
\begin{equation}\label{R-transform-sum}
\mathcal{R}_{A+B}(y)=\mathcal{R}_A(y)+\mathcal{R}_B(y),
\end{equation}
and the scaling rule of a random matrix $A$ with a scalar $\mu$,
\begin{equation}\label{R-transform-scaling}
\mathcal{R}_{\mu A}(y)=\mu \mathcal{R}_A(\mu y),
\end{equation}
we have
\begin{equation}
\mathcal{R}_{Y_L}(y)=-e^{i\pi/2}y^{-1/2},
\end{equation}
too. We have numerically illustrated this for $L=1,2,3,4$ in Fig.~\ref{fig:rhomacro-alp05-bet1}.

\begin{figure}[t!]
\begin{center}
\includegraphics[width=0.8\textwidth]{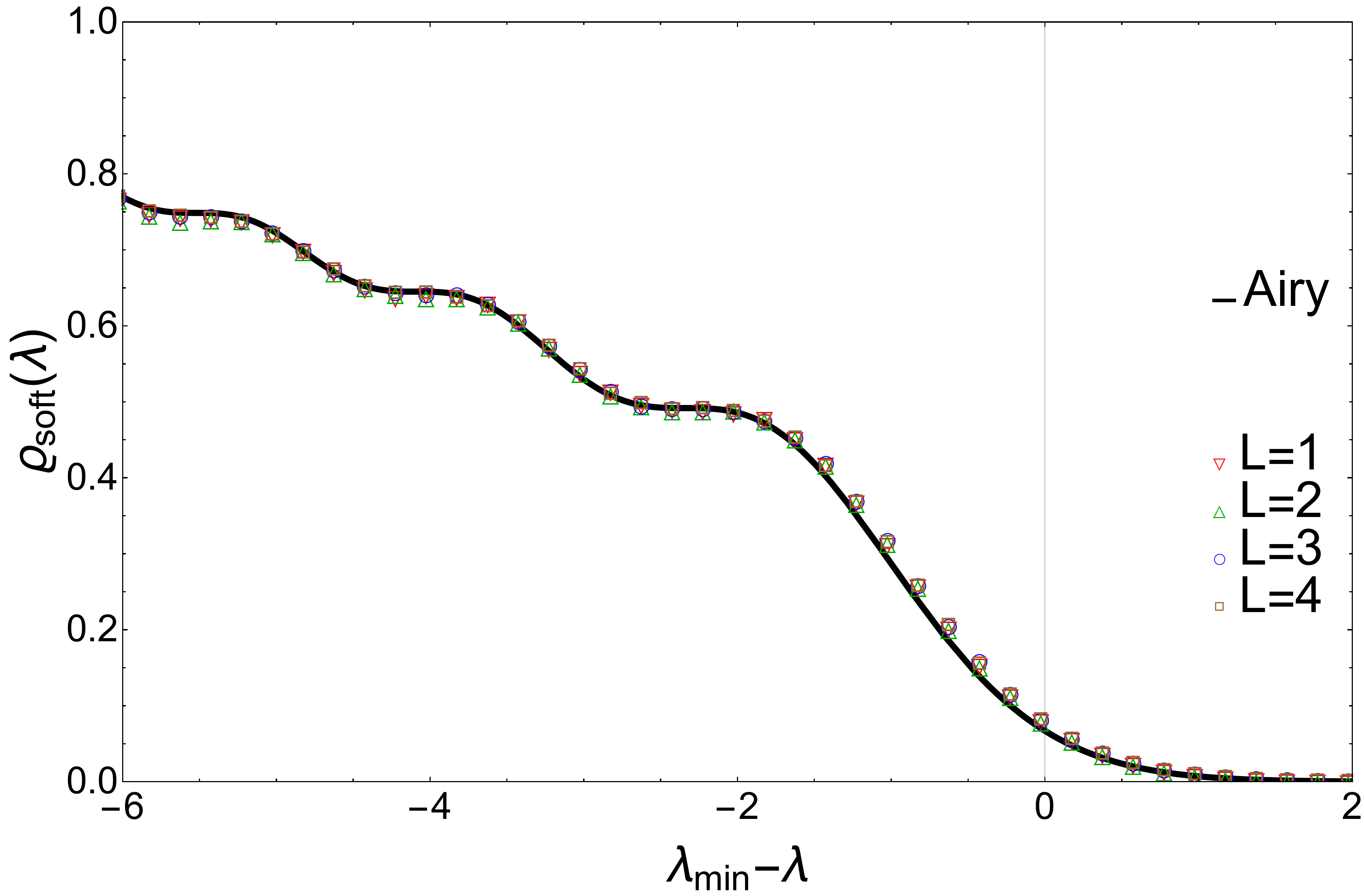}
\caption{Microscopic level density at the soft edge for the Monte Carlo simulations (coloured symbols) of the random matrix sums~\eref{sum-random} and the analytical prediction~\eref{soft-edge.micro} (solid black curve). We have employed the same configurations generated for Fig.~\ref{fig:rhomacro-alp05-bet1}. The bin size is this time $0.2$.}
\label{fig:rhosoft-alp05-bet1}
\end{center}
\end{figure}

Other statistics that can be checked to be stable when performing the sum~\eref{sum-random} are those in the bulk and at the soft-edge. In Ref.~\cite{LWZ2016}, those have been proven to be those shared with the GUE. For instance, the soft edge lies at $\lambda_{\min}=1/4$ in the macroscopic scaling for any $L\in\mathbb{N}$. Thus, we should find the microscopic level density~\cite{Mehtabook,Peterbook}
\begin{eqnarray}
\rho_{\rm Airy}(\lambda)&=&\lim_{N\to\infty}\left\langle\tr\delta\left(\lambda\id_N-\frac{N^{2/3}}{2^{1/3}}\left[\id_N-\frac{4N}{L^2}\sum_{j=1}^LX_j^\dagger X_j\right]\right)\right\rangle\nonumber\\
&=&[{\rm Ai}'(\lambda)]^2-\lambda[{\rm Ai}(\lambda)]^2
\label{soft-edge.micro}
\end{eqnarray}
with the Airy function ${\rm Ai}(x)$ and its derivative ${\rm Ai}'(x)$. This stability has been corroborated in Fig.~\ref{fig:rhosoft-alp05-bet1} with the help of Monte Carlo simulations.

The question is what are the eigenvalue statistics in the tail? For $L=1$ it is rather trivial because $(X_j^\dagger X_j)^{-1}$ is an ordinary Laguerre ensemble that exhibits a hard edge with the microscopic level density~\cite{Mehtabook,Peterbook} of the Bessel kernel,
\begin{eqnarray}
\rho_{\rm Bessel}(\lambda)&=&\lim_{N\to\infty}\left\langle\tr\delta\left(\lambda\id_N-\frac{2\sqrt{N}}{\pi}(X_1^\dagger X_1)^{-1/2}\right)\right\rangle\nonumber\\
&=&\frac{\pi^2}{2}\lambda[J_0^2(\pi \lambda)+J_1^2(\pi \lambda)].
\label{hard-edge.micro}
\end{eqnarray}
We have used the Bessel function of the first kind $J_\nu(x)$. We also note that we have unfolded in such a way that the asymptotic of the level density for large $\lambda$ is $\rho_{\rm Bessel}(\lambda)\to1$, meaning the mean level spacing is approximately one.

From~\eref{hard-edge.micro} we can read off the microscopic level density in the tail which we `baptise' the inverse Bessel statistics,
\begin{eqnarray}
\rho_{\rm invB}(\lambda)&=&\frac{1}{\lambda^2}\rho_{\rm Bessel}(\lambda^{-1})=\frac{\pi^2}{2\lambda^3}[J_0^2(\pi \lambda^{-1})+J_1^2(\pi \lambda^{-1})].
\label{tail.micro}
\end{eqnarray}
We note that the tail of the largest eigenvalues is decaying slightly stronger than the macroscopic level density, namely $\rho_{\rm invB}(\lambda)\propto \lambda^{-3}$ in contrast to $\rho_{X_1^\dagger X_1}(\lambda)\propto\lambda^{-3/2}$. One reason is that the unfolding involves taking the square root of the eigenvalues. Yet, this does not completely explain everything. The largest eigenvalues are far in the tail and the different tail behaviours lie on a different scale than the heavy-tail of the macroscopic level density. This is known already for the hard edge where the macroscopic level density can have a totally different behaviour (for instance it goes to zero instead of diverging as a square root singularity) than the microscopic ones due to the very different scales.

\begin{figure}[t!]
\begin{center}
\includegraphics[width=0.8\textwidth]{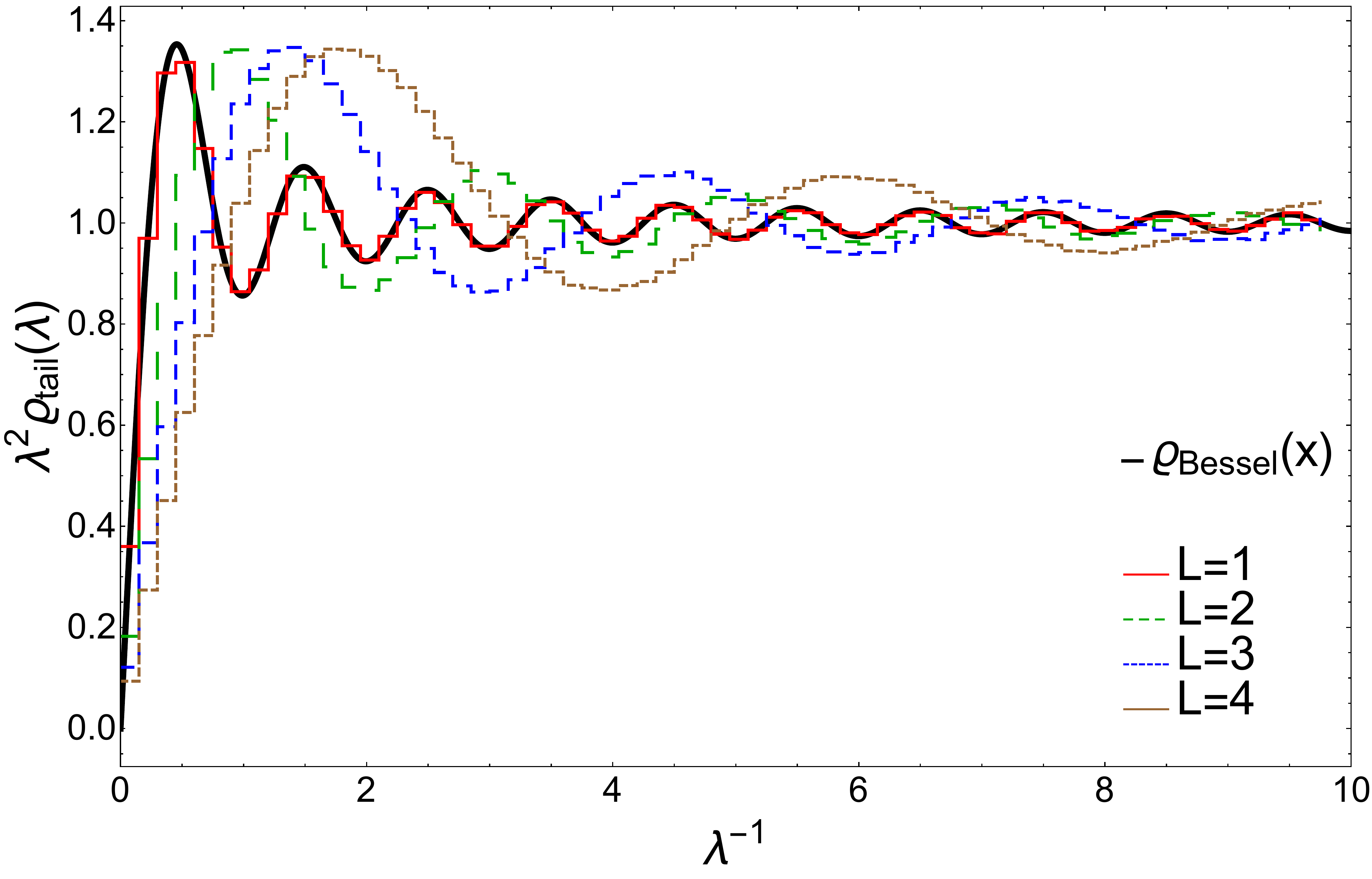}
\caption{The microscopic level density of the Monte Carlo simulations of Figs.~\ref{fig:rhomacro-alp05-bet1} and~\ref{fig:rhosoft-alp05-bet1} (coloured histograms) compared to the inverse Bessel statistic result~\eref{tail.micro} (black solid curve). The bin size is this time $0.2$. We underline that we have employed the unfolded scale meaning that we have inverted the spectrum so that the largest eigenvalues are those closest to the origin.}
\label{fig:rhotail-alp05-bet1}
\end{center}
\end{figure}

\begin{figure}[t!]
\begin{center}
\includegraphics[width=\textwidth]{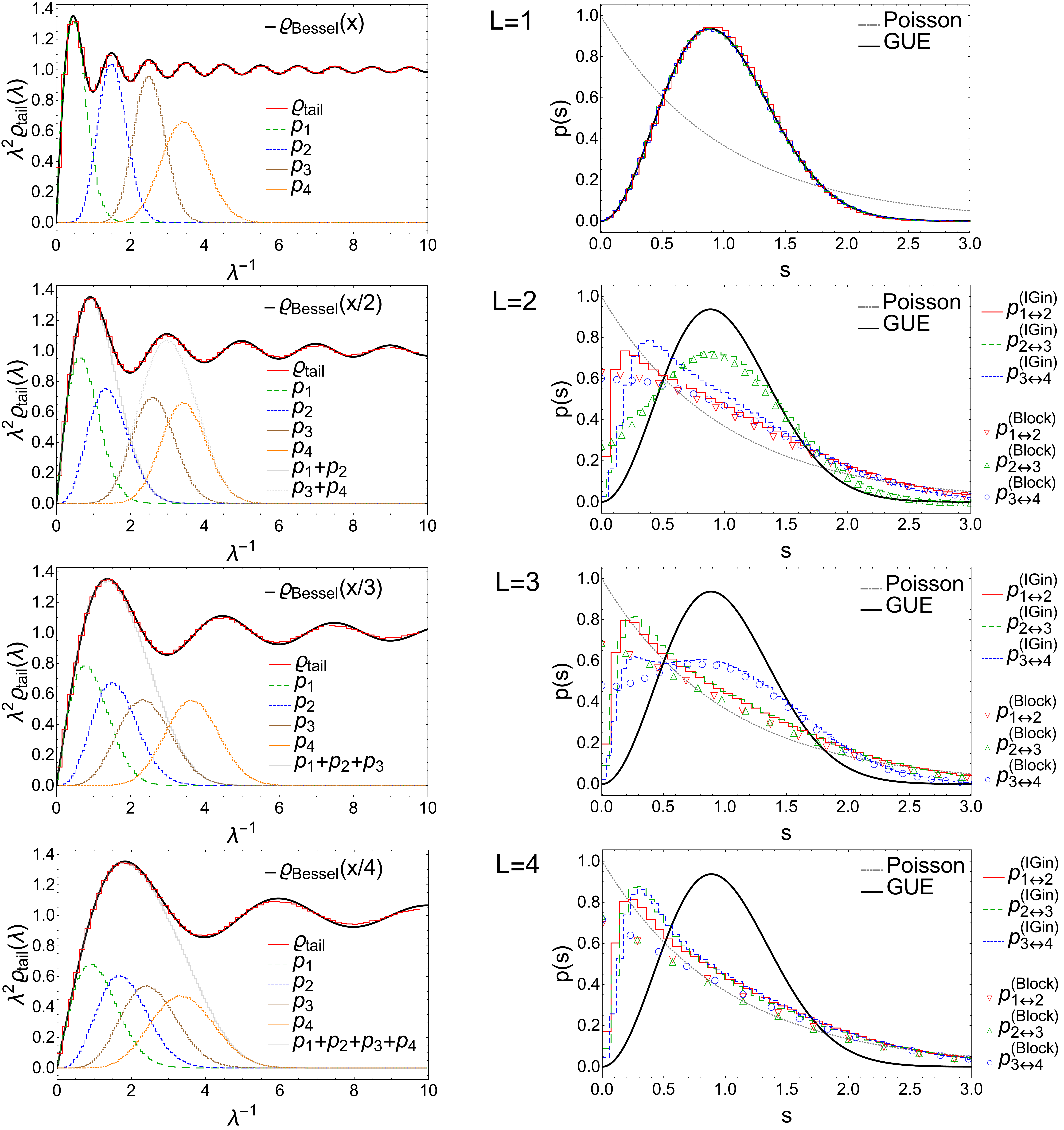}
\caption{{\bf Left plots:} distributions of the individual eigenvalues (dashed coloured histograms, bin size $0.1$, only Monte Carlo simulated) compared with the microscopic tail level densities~\eref{tail.micro.L} in the unfolded (inverted) scale (red solid histograms for the simulation with bin size $0.2$ and black solid curve for the analytical result) and with the sum of the distribution for those eigenvalues that cluster to one of the peaks. The configurations are the same as those of Figs.~\ref{fig:rhomacro-alp05-bet1},~\ref{fig:rhosoft-alp05-bet1}, and~\ref{fig:rhotail-alp05-bet1}. {\bf Right plots:} unfolded level spacing distributions of the three largest eigenvalues for the random matrix sum~\eref{sum-random} (coloured histograms) and for the direct sum~\eref{direct-sum-random} (coloured symbols), both with a bin size $0.1$. For reference we have also drawn the spacing distributions of the Poisson statistics~\eref{Poisson-dist} and the Wigner surmise~\eref{GUE-surmise-dist} of the GUE.}
\label{fig:rhotailind-alp05-bet1}
\end{center}
\end{figure}

We would like investigate whether the tail statistics are also stable and particularly if Eq.~\eref{tail.micro} still holds for the sum~\eref{sum-random} for any $L\in\mathbb{N}$. Unfortunately and quite surprisingly this is not the case. Numerically we have observed that the microscopic level density in the tail follows the law
\begin{eqnarray}
\fl\rho_{\rm invB}^{(Y_L)}(\lambda)&=&L^2\rho_{\rm invB}(L\lambda)=\frac{1}{\lambda^2}\rho_{\rm Bessel}((L\lambda)^{-1})=\frac{\pi^2}{2L\lambda^3}[J_0^2(\pi (L\lambda)^{-1})+J_1^2(\pi (L\lambda)^{-1})],\nonumber\\
\fl&&
\label{tail.micro.L}
\end{eqnarray}
see Fig.~\ref{fig:rhotail-alp05-bet1} and left plots of Fig.~\ref{fig:rhotailind-alp05-bet1}. Note that this is not a simple rescaling for the level density as then we would have multiplied the density by $L$ and not $L^2$ resulting from the Jacobian. Therefore there has to be a different and perhaps more fundamental change in the spectral statistics.

What is actually going on in the tail? To analyse this first numerically we have measured the distributions of the four largest eigenvalues and the three level spacing distributions between these four eigenvalues of $Y_L$ for $L=1,2,3,4$, see Fig.~\ref{fig:rhotailind-alp05-bet1}. Each single maximum of the microscopic Bessel level density~\eref{hard-edge.micro} and thus the inverse Bessel result~\eref{tail.micro.L} is now described by $L$ eigenvalues and not just a single one. The level spacing distribution indeed corroborates that the eigenvalues corresponding to one maximum have a diminished level repulsion, see to the right plots of Fig.~\ref{fig:rhotailind-alp05-bet1}.

Our interpretation is that in the large $N$-limit of $Y_L$ the eigenvalue tail statistics asymptotes to the statistics of the direct sum
\begin{equation}\label{direct-sum-random}
\widehat{Y}_L=\bigoplus_{j=1}^LX_j^\dagger X_j=\left[\begin{array}{ccc} X_1^\dagger X_1 & & 0 \\ & \ddots & \\ 0 & & X_L^\dagger X_L\end{array}\right].
\end{equation}
We have also simulated these random matrices and measured the same level spacing distributions of the largest eigenvalues for comparison. In the right plots of Fig.~\ref{fig:rhotailind-alp05-bet1} we see indeed similarities such that our point is substantiated. The deviations can be understood as residual level repulsions which are suppressed onto very small scales. Therefore we expect a convergence to the statistics of~\eref{direct-sum-random} though it will not be uniform about very small spacings. For reference we have added the level spacing distribution for the Poisson statistics (statistically independent eigenvalues).
\begin{equation}\label{Poisson-dist}
p_{\rm Poisson}(s)=e^{-s}
\end{equation}
and the Wigner surmise for the GUE
\begin{equation}\label{GUE-surmise-dist}
p_{\rm Poisson}(s)=\frac{32}{\pi^2}x^2\exp\left[-\frac{4}{\pi}x^2\right],
\end{equation}
which is pretty close to the universal bulk distribution~\cite{DH1990} as well as to the level spacing distributions at the hard edge~\cite{AGK2021}.

In Sec.~\ref{sec:analytical} we have derived an analytical argument based on the supersymmetry method that verifies our understanding. The question that remains is if this observation is true in general and whether it is perhaps governed by a more universal behaviour. This will be discussed later, see Sec.~\ref{sec:conclusio}.

\section{Analytical Corroborations}\label{sec:analytical}

In Sec.~\ref{sec:tail.scale} we define and discuss the random matrix ensembles namely the sum of products of inverse Wishart-Laguerre matrices that we study with the supersymmetry method in Sec.~\ref{sec:SUSY}. After we have derived the corresponding supermatrix integral we carry out the large $N$-limit in Sec.~\ref{sec:asymptotic}. Finally in Sec.~\ref{sec:critical} we look for the critical scale where the spectral statistics of this ensemble changes from stable (meaning the sum exhibits the same statistics as each matrix in the sum) to unstable.

\subsection{Matrix Model and the Scaling of the Tail Statistics}\label{sec:tail.scale}

After we have seen a glimpse of the peculiarities of the spectral statistics in heavy tails we will show in this section that first the proper interpretation is indeed the one we have concluded with in the last section and secondly that it works for more general ensembles. For this reason we consider a sum of product matrices of the form
\begin{equation}\label{XLM}
X_l^{(M)}=X_{l,M}X_{l,M-1}\cdots X_{l,1},\quad {\rm for}\ l=1,\ldots,L,
\end{equation}
where each $X_{l,m}\in\mathbb{C}^{N\times N}$ is independently drawn from~\eref{Inv-Gin}. Then the sum is given by
\begin{equation}\label{YLM}
Y_L^{(M)}=\frac{1}{L^{M+1}}\sum_{l=1}^L(X_l^{(M)})^\dagger X_l^{(M)}.
\end{equation}
Comparison with~\eref{sum-random} shows that the stability exponent will be $\alpha=1/(M+1)$. Certainly it has been shown that the macroscopic level density of $Y_1^{(M)}$,
\begin{equation}\label{MP.M-law}
\rho_{Y_1^{(M)}}(\lambda)=\lim_{N\to\infty}\left\langle\frac{1}{N}\tr\delta\left(\lambda\id_N-N^MY_1^{(M)}\right)\right\rangle,
\end{equation}
 is stable under free convolution, too, see~\cite{BPB1999,APA2009}. The reason is the $\mathcal{S}$-transform, another transform in free probability indirectly defined via the $\mathcal{R}$-transform~\cite{Speicher2011}
 \begin{equation}\label{S-transform}
 \mathcal{R}(y)=\frac{1}{\mathcal{S}[y\mathcal{R}(y)]}\qquad\Leftrightarrow\qquad \mathcal{S}(\chi)=\frac{1}{\mathcal{R}[\chi \mathcal{S}(\chi)]}.
 \end{equation}
 It has the property~\cite{Speicher2011}
 \begin{equation}
 \mathcal{S}_{Y_1^{(M)}}(\chi)= \mathcal{S}_{X_{1,M}^\dagger Y_1^{(M-1)}X_{1,M}}(\chi)= \mathcal{S
}_{Y_1^{(M-1)}}(\chi) \mathcal{S}_{X_{1,M}^\dagger X_{1,M}}(\chi)
 \end{equation}
 as $X_{1,M}$ and $Y_1^{(M)}$ are asymptotically free. In our case we have
 \begin{eqnarray}
 \fl \mathcal{S}_{X_{l,m}^\dagger X_{l,m}}(\chi)=-\chi\ &\Rightarrow&\  \mathcal{S}_{Y_l^{(M)}}(\chi)=(-\chi)^M\ \Rightarrow\  \mathcal{R}_{Y_l^{(M)}}(y)=-e^{i\pi/(M+1)}y^{-M/(M+1)}.\nonumber\\
 \fl\label{R-S-transform-M}
 \end{eqnarray}
 This result reflects the stability when considering the classification in~\cite[Appendix A]{BPB1999}.
 
  The macroscopic level density of the inverse of $Y_1^{(M)}$ is given in terms of the Meijer G-function~\cite{PZ2011}
 \begin{eqnarray}\label{macro-dens.M}
 \fl\rho_{(Y_1^{(M)})^{-1}}(\lambda)&=&\frac{1}{\sqrt{2\pi}}\frac{M^{M-3/2}}{(M+1)^{M+1/2}}\\
 \fl&&\times G^{M,0}_{M,M}\left(\left.\begin{array}{c} \{(1+j-M)/M\}_{j=1,\ldots,M} \\ \{(j-1-M)/(M+1)\}_{j=1,\ldots,M} \end{array}\right|\frac{M^M}{(M+1)^{M+1}}\lambda\right).\nonumber
 \end{eqnarray}
 The Meijer G-function~\cite{ASbook} is essentially an inverse Mellin transform of ratios of Gamma functions, i.e.
 \begin{eqnarray}\label{Meijer-G}
 &&G^{p,q}_{m,n}\left(\left.\begin{array}{c} a_1,\ldots,a_n;b_1,\ldots,b_p \\ c_1,\ldots,c_m;d_1,\ldots,d_q \end{array}\right|z\lambda\right)\\
 &=&\int_{\mathcal{C}}\frac{\left(\prod_{j=1}^m\Gamma[c_j+s]\right)\left(\prod_{j=1}^n\Gamma[1-a_j-s]\right)}{\left(\prod_{j=1}^p\Gamma[b_j+s]\right)\left(\prod_{j=1}^q\Gamma[1-d_j+s]\right)}z^{-s}\frac{ds}{2\pi i},\nonumber
 \end{eqnarray}
 where the contour $\mathcal{C}$ starts at $-i\infty$ and finishes at $+i\infty$ while having the poles of $\Gamma[c_j+s]$ on the left side of the path and $\Gamma[1-a_j-s]$ are on the right side.
 The distribution~\eref{macro-dens.M} is called the Fuss-Catalan distribution since it has the Fuss-Catalan numbers~\cite{PZ2011},
 \begin{equation}
FC_M(n)=\frac{\Gamma[(M+1)n-M]}{\Gamma[M n-M+2]\Gamma[n]}\quad {\rm with}\ n\in\mathbb{N}_0,
\end{equation}
as its moments and it has a support on $\lambda\in[0,(M+1)^{M+1}/M^M]$. Additionally, its behaviour at the origin diverges like
\begin{equation}
 \rho_{(Y_1^{(M)})^{-1}}(\lambda)\approx \frac{\lambda^{-M/(1+M)}}{\Gamma[(M+2)/(M+1)]\Gamma[M/(M+1)]}\quad {\rm for}\ \lambda\ll1.
\end{equation}
this implies that the tail behaviour of the matrix $Y_1^{(M)}$ will be
\begin{equation}
 \fl\rho_{Y_1^{(M)}}(\lambda)= \lambda^{-2}\rho_{(Y_1^{(M)})^{-1}}(\lambda^{-1})\approx \frac{\lambda^{M/(1+M)-2}}{\Gamma[(M+2)/(M+1)]\Gamma[M/(M+1)]}\quad {\rm for}\ \lambda\gg1.
\end{equation}
where we can again read off the stability exponent $\alpha=1/(M+1)$ which is consistent with the other discussion.

As a result from the above discussion, the scaling of the largest eigenvalues of $Y_1^{(M)}$ as well as $Y_L^{(M)}$ will be $N$ regardless of $M$. This scaling can be obtained by combining $N\rho_{Y_1^{(M)}}(\lambda)d\lambda\propto d(N\lambda^{-1/(M+1)})$ for $\lambda\gg1$ and Eq.~\eref{MP.M-law}. We need this scale to properly unfold the spectrum as well as to find the largest eigenvalues. For instance, the unfolded microscopic tail level density of $Y_1^{(M)}$ is given by the so-called Meijer G-kernel result~\cite[Theorem 5.3]{KZ2014} ($\nu_j=0$ for all $j$) of the hard edge microscopic level density of $(Y_1^{(M)})^{-1}$ which is
\begin{eqnarray}
\fl\rho_{\rm Meijer G}^{(M)}(\lambda)&=&\lim_{N\to\infty}\left\langle\tr\delta\left(\lambda\id_N-\frac{N^{1/(M+1)}}{c_M}(Y_1^{(M)})^{-1/(M+1)}\right)\right\rangle\nonumber\\
\fl&=&(M+1)c_M^{M+1}\lambda^M\int_0^1dt G^{0,M+1}_{1,0}\left(\left.\begin{array}{c} -;- \\ 0;0,\ldots,0 \end{array}\right|t(c_M\lambda)^{M+1}\right)\nonumber\\
\fl&&\times G^{0,M+1}_{M,0}\left(\left.\begin{array}{c} -;- \\ 0,\ldots,0;0 \end{array}\right|t(c_M\lambda)^{M+1}\right)
\label{Meijer-G.micro}
\end{eqnarray}
with
\begin{equation}\label{unfold.const}
c_M=\Gamma\left[\frac{M+2}{M+1}\right]\Gamma\left[\frac{M}{M+1}\right]
\end{equation}
the proper unfolding constant such that $\rho_{\rm Meijer G}^{(M)}(\lambda)\approx\lambda$ for $\lambda\gg1$.
For $M=2$, this formula reduces to~\eref{hard-edge.micro}. Due to the proper unfolding  the spectrum becomes the half sided picket fence spectrum for $M\to\infty$~\cite{ABK2014,ABK2019,LWW2018,ABK2020}
\begin{eqnarray}
\lim_{M\to\infty}\rho_{\rm Meijer G}^{(M)}(\lambda)&=&\sum_{j=1}^\infty\delta(\lambda-j+0.5).
\label{picket.fence}
\end{eqnarray}
The shift by $0.5$ reflects the level repulsion from the origin and has thus a strong resemblance to the spectrum of the quantum harmonic oscillator.

The microscopic tail level density is given then by 
\begin{eqnarray}
 \rho_{\rm invMG}^{(M)}(\lambda)&=&\frac{1}{\lambda^2}\rho_{\rm Meijer G}^{(M)}(\lambda^{-1}).
\label{invMeijer-G.micro}
\end{eqnarray}
This is the one that can be expected when studying the matrix $Y_1^{(M)}$. For the sum of $L$ copies of matrices, meaning $Y_L^{(M)}$, we will find that the averaged spectrum behaves as if we would have directly summed these random matrices, cf., subsection~\ref{sec:asymptotic}.

\subsection{Supersymmetry Method}\label{sec:SUSY}

Instead of computing the level density or more generally the $k$-point correlation functions of $Y_L^{(M)}$ we consider the partition function
\begin{equation}\label{partition.def}
Z_{Y_L^{(M)}}^{(k,N)}(\kappa)=\left\langle\prod_{j=1}^k\frac{\det(Y_L^{(M)}-\kappa_{\F,j})}{\det(Y_L^{(M)}-\kappa_{\B,j})}\right\rangle
\end{equation}
with $\kappa_{\F,j},\kappa_{\B,j}\in\mathbb{C}$ and $\kappa_{\B,j}\notin[0,+\infty[$ for all $j=1,\ldots,k$. It is a well-known fact these kinds of partition functions can generate several different quantities such as the $k$-point correlation functions and specifically the level density, see Refs.~\cite{Mehtabook,Efetovbook,Zirnbauer2006,Guhr2011}. The variables $\kappa_{\F,j}$ and $\kappa_{\B,j}$ usually contain the spectral variables that correspond to the eigenvalue statistics, regularisations such as an imaginary shift away from the positive real axis, as well as some source variables that can be expanded and can create Green's functions.
When arranging the $\kappa_{\F,j}$ and $\kappa_{\B,j}$ in the form of a diagonal supermatrix $\kappa=\diag(\kappa_{\B,1},\ldots,\kappa_{\B,k};\kappa_{\F,1},\ldots,\kappa_{\F,k})$, we can write the average in terms of a superdeterminant
\begin{equation}\label{partition.sdet}
Z_{Y_L^{(M)}}^{(k,N)}(\kappa)=\left\langle\Sdet(Y_L^{(M)}\otimes\id_{k|k}-\id_N\otimes\kappa)^{-1}\right\rangle.
\end{equation}

For an  introduction to superalgebra and superanalysis we refer to~\cite{Berezinbook}. As the conventions slightly vary in the literature, we briefly summarise ours in the present work. A $(p_1|q_1)\times(p_2|q_2)$ supermatrix $A$ can be arranged into four blocks
\begin{equation}
A=\left[\begin{array}{cc} A_{\B\B} & A_{\B\F} \\ A_{\F\B} & A_{\F\F} \end{array}\right]
\end{equation}
with the Boson-Boson ($A_{\B\B}:\ p_1\times p_2$) and the Fermion-Fermion ($A_{\F\F}:\ q_1\times q_2$) blocks containing only commuting variables and the Boson-Fermion ($A_{\B\F}:\ p_1\times q_2$) and the Fermion-Boson ($A_{\F\B}:\ q_1\times p_2$) comprising of only anti-commuting variables. Then for $p_1=p_2=p$ and $q_1=q_2=q$, the supertrace and the superdeterminant are given by
\begin{equation}\label{superalgebra}
\fl\Str A=\tr A_{\B\B}-\tr A_{\F\F}\qquad {\rm and}\qquad \Sdet A=\frac{\det(A_{\B\B}-A_{\B\F}A_{\F\F}^{-1}A_{\F\B})}{\det A_{\F\F}},
\end{equation}
respectively. Each commuting variable consists of a numerical part, which can be real or complex, and a nilpotent one which is an even polynomial of the underlying Grassmann variables $\{\eta_j\}$ (algebraic basis of the anti-commuting variables), while an anti-commuting variable only contains a nilpotent part which is an odd polynomial of the Grassmann variables. The integral over a Grassmann variable is defined by the two axiomatic identities
\begin{equation}\label{superanalysis}
\int d\eta=0\qquad{\rm and}\qquad\int \eta d\eta=1.
\end{equation}
These two equalities are enough to fix the integration over Grassmann variables as any function of Grassmann variables is understood as a finite Taylor series due to the nilpotence of the Grassmann variables, i.e., $\eta_j^2=0$.

Coming back to~\eref{partition.sdet}, we can readily generalise the partition function to an arbitrary $(k|k)\times(k|k)$ supermatrix $\kappa$ as long as the numerical part of the eigenvalues of the Boson-Boson block $\kappa_{\B\B}$ do not lie on the positive real line or the origin.

The idea of the supersymmetry method in random matrix theory is to map the average over an ordinary random matrix to an average over a supermatrix whose dimension is independent of $N$. In our case we essentially have products of matrices of the form $WW^\dagger$. In Refs.~\cite{KKG2014,Kieburg2015}, one of the present authors has introduced a short cut called the supersymmetric projection formula that essentially states for a random matrix $W\in\mathbb{C}^{N\times N}$ that is distributed along a unitarily invariant density $P(WW^\dagger)=P(VWW^\dagger V^\dagger)$ (for all $V\in\U(N)$ and $W\in\mathbb{C}^{N\times N}$), we can find a superfunction $Q(U)$ for a $(k|k)\times(k|k)$ supermatrix $U$ such that
\begin{equation}\label{duality.SUSY}
\left\langle\Sdet(WW^\dagger\otimes\id_{k|k}+\widehat{\kappa})^{-1}\right\rangle=\left\langle\Sdet(\id_N\otimes U+\widehat{\kappa})^{-1}\right\rangle.
\end{equation}
Here, $\widehat{\kappa}$ can be a much larger supermatrix of dimensions $(kN|kN)\times(kN|kN)$. We underline that on the left hand side we average over the ordinary random matrix $W$ while on the right hand side we average over the supermatrix $U$.

The supermatrix $U$ is in the current situation relatively simple, namely its Boson-Boson block is a positive definite Hermitian matrix $U_{\B\B}=U_{\B\B}^\dagger\in\Herm_+(k)$ with no Grassmann variables and its Fermion-Fermion block is a unitary matrix $U_{\F\F}\in\U(k)$ also containing no Grassmann variables. The Boson-Fermion and Fermion-Boson blocks only comprise independent Grassmann variables with no further symmetries. Therefore, the supermatrix space described by $U$ is the supersymmetric coset $\Herm_\odot(k|k)={\rm Gl}(k|k)/\U(k|k)$, see~\cite{Zirnbauer1996}. The superfunction $Q(U)$ is given via the supersymmetric projection formula~\cite{KKG2014,Kieburg2015} times the measure $\Sdet U^{N}d[U]$ on $\Herm_\odot(k|k)$,
\begin{equation}\label{SUSY.projection}
\fl Q(U)=\int_{\mathbb{C}^{N\times N}} d[W]\int_{\mathbb{C}^{N\times(k|k)}} d[\widetilde{W}] \widetilde{P}\left(\left[\begin{array}{cc}  WW^\dagger+\widetilde{W}\widetilde{W}^\dagger & \widetilde{W} \\ U\widetilde{W}^\dagger & U\end{array}\right]\right),
\end{equation}
 where $d[U]$ is the product of the differentials of all supermatrix elements. The supermatrix $\widetilde{W}$ is a rectangular matrix where its first $k$ columns are ordinary $N$ dimensional complex vectors and the last $k$ columns are $N$ dimensional vectors with independent complex Grassmann variables as their entries, this set has been denoted by $\mathbb{C}^{N\times(k|k)}$. The superfunction $\widetilde{P}$ is a supersymmetric extension of $P$ that satisfies
\begin{equation}
P(V V^\dagger)=\widetilde{P}(V^\dagger V),\ {\rm for\ any}\ V\in\mathbb{C}^{N\times(N+k|k)}.
\end{equation}
Such a supersymmetric extension is commonly not unique but there are usually natural choices as we will see below for the present situation of the inverse Ginibre matrices.

We approach the average~\eref{partition.sdet} inductively by writing $Y_L^{(M)}$ as a sum and a product of inverse Ginibre matrices, 
\begin{eqnarray}
\fl Z_{Y_L^{(M)}}^{(k,N)}(\kappa)&=&\left\langle\Sdet({X_L^{(M)}}^\dagger X_L^{(M)}\otimes\id_{k|k}+L^{M+1}Y_{L-1}^{(M)}\otimes\id_{k|k}-L^{M+1}\id_N\otimes\kappa)^{-1}\right\rangle\nonumber\\
\fl&=&\left\langle\Sdet(X_{L,M}^\dagger X_{L,M}\otimes\id_{k|k}+\widehat{\kappa}_{L,M})^{-1}\right\rangle\label{partition.SUSY.1}
\end{eqnarray}
with
\begin{equation}
\fl\widehat{\kappa}_{L,M}=L^{M+1}\left[({X_L^{(M-1)}}^\dagger)^{-1}Y_{L-1}^{(M)}(X_L^{(M-1)})^{-1}\otimes\id_{k|k}-({X_L^{(M-1)}}^\dagger X_L^{(M-1)})^{-1}\otimes\kappa\right].
\end{equation}
For the rearrangement of the matrices inside the superdeterminant we have employed the identities
\begin{equation}
\Sdet(AB)=\Sdet(A)\Sdet(B)\quad {\rm and}\quad \Sdet(H\otimes\id_{k|k})=1
\end{equation}
for any two square supermatrices $A$ and $B$ and any ordinary square matrix $H$.

Next, we apply~\eref{duality.SUSY} to~\eref{partition.SUSY.1} and obtain
\begin{eqnarray}\label{partition.SUSY.2}
\fl Z_{Y_L^{(M)}}^{(k,N)}(\kappa)&=&\left\langle\Sdet(\id_N\otimes U_{L,M}+\widehat{\kappa})^{-1}\right\rangle\\
\fl&=&\left\langle\Sdet({X_L^{(M-1)}}^\dagger X_L^{(M-1)}\otimes U_{L,M}+L^{M+1}[Y_{L-1}^{(M)}\otimes \id_{k|k}-\id_N\otimes\kappa])^{-1}\right\rangle.\nonumber
\end{eqnarray}
We repeat this procedure until each integration over the random matrix $X_{L,j}$ is transferred into an integration over a supermatrix $U_{L,j}$. The only thing what changes is the supermatrix
\begin{eqnarray}
\fl\widehat{\kappa}_{L,j}&=&L^{M+1}\biggl[({X_L^{(j-1)}}^\dagger)^{-1}Y_{L-1}^{(M)}(X_L^{(j-1)})^{-1}\otimes(U_{L,j+1}\cdots U_{L,M})^{-1}\\
\fl&&-({X_L^{(j-1)}}^\dagger X_L^{(j-1)})^{-1}\otimes\kappa(U_{L,j+1}\cdots U_{L,M})^{-1}\biggl].
\end{eqnarray}
This eventually leads to
\begin{eqnarray}\label{partition.SUSY.3}
\fl Z_{Y_L^{(M)}}^{(k,N)}(\kappa)&=&\left\langle\Sdet(L^{-M-1}\id_N\otimes U_{L}^{(M)}+Y_{L-1}^{(M)}\otimes \id_{k|k}-\id_N\otimes\kappa)^{-1}\right\rangle
\end{eqnarray}
with
\begin{equation}
U_{L}^{(M)}=U_{L,1}\cdots U_{L,M}.
\end{equation}

The average over $Y_{L-1}^{(M)}$ looks now like the one over $Y_{L}^{(M)}$. Therefore, we can proceed as before and find eventually
\begin{eqnarray}\label{partition.SUSY.4}
 Z_{Y_L^{(M)}}^{(k,N)}(\kappa)&=&\left\langle\Sdet(V_L^{(M)}-\kappa)^{-N}\right\rangle
\end{eqnarray}
with
\begin{equation}\label{SUSY.LM}
V_L^{(M)}=\frac{1}{L^{M+1}}\sum_{j=1}^LU_j^{(M)}\ {\rm and}\ U_j^{(M)}=U_{j,1}\cdots U_{j,M}
\end{equation}
in analogy to~\eref{XLM} and~\eref{YLM}. 

The remaining ingredient to be calculated is the superfunction $Q$ for the inverse Ginibre ensemble. For this purpose, it helps to know the corresponding superfunction for the Ginibre ensemble that has been computed with the projection formula in~\cite{Kieburg2015}, which follows from the Gaussian structure
\begin{equation}
\exp[-\tr VV^\dagger]=\exp[-\Str V^\dagger V],\ {\rm for\ any}\ V\in\mathbb{C}^{N\times(N+k|k)},
\end{equation}
thence,
\begin{eqnarray}
\fl Q_{\rm Gin}(U_{\rm Gin})&=&\int_{\mathbb{C}^{N\times N}} d[W]\int_{\mathbb{C}^{N\times(k|k)}} d[\widetilde{W}] \exp\left(-\Str\left[\begin{array}{cc}  WW^\dagger+\widetilde{W}\widetilde{W}^\dagger & \widetilde{W} \\ U_{\rm Gin}\widetilde{W}^\dagger & U_{\rm Gin}\end{array}\right]\right)\nonumber\\
\fl&\propto&\exp[-\Str U_{\rm Gin}].
\end{eqnarray}
We can make use of it by noticing that the averages~\eref{duality.SUSY} between Ginibre and inverse Ginibre are related,
\begin{eqnarray}
\fl\left\langle\Sdet(WW^\dagger\otimes\id_{k|k}+\widehat{\kappa})^{-1}\right\rangle&=&\left\langle\Sdet((WW^\dagger)^{-1}\otimes\id_{k|k}+\widehat{\kappa}^{-1})^{-1}\Sdet\widehat{\kappa}^{-1}\right\rangle\\
\fl&=&\left\langle\Sdet(\id_N\otimes U_{\rm Gin}+\widehat{\kappa}^{-1})^{-1}\Sdet\widehat{\kappa}^{-1}\right\rangle\nonumber\\
\fl&=&\left\langle(\Sdet U_{\rm Gin})^{-N}\Sdet(\id_N\otimes U_{\rm Gin}^{-1}+\widehat{\kappa})^{-1}\right\rangle.\nonumber
\end{eqnarray}
Simple comparison with the general duality~\eref{duality.SUSY} yields the identification that each supermatrices $U_{l,m}$ in~\eref{partition.SUSY.4} is a copy of $U_{\rm Gin}^{-1}$ which we coin $V_{l,m}$. 

Summarising, the partition function takes the form
\begin{eqnarray}\label{partition.final}
\fl  Z_{Y_L^{(M)}}^{(k,N)}(\kappa)&=&\frac{\int_{\Herm_\odot^{ML}(k|k)}\Sdet(V_L^{(M)}-\kappa)^{-N}\prod_{m=1}^M\prod_{l=1}^Le^{-\Str V_{l,m}}d[V_{l,m}]}{(\int_{\Herm_\odot(k|k)}(\Sdet V)^{N}e^{-\Str V}d[V])^{LM}}
\end{eqnarray}
with
\begin{equation}\label{SUSY.LM.b}
V_L^{(M)}=\frac{1}{L^{M+1}}\sum_{j=1}^LV_{j,1}^{-1}\cdots V_{j,M}^{-1}.
\end{equation}
The normalisation $Z_{Y_L^{(M)}}^{(k,N)}(z\id_{k|k})=1$ with an arbitrary complex $z\notin[0,+\infty[$ follows from the Wegner integration theorems~\cite{PS1979,Wegner1983,Efetov1983,Constantinescu1988,CdeG1989,KKG2009,Kieburg2011} for supergroup invariant integrands. These theorems are essentially multidimensional Cauchy-like identities which tell us that the integral is essentially the integrand at $V\propto\id_{k|k}$ times an integrand independent constant. The proportionality constant cancels in the invariants like the supertrace or the superdeterminant as can be readily checked with~\eref{superalgebra}. Therefore, the denominator in~\eref{partition.final} is only an $N$ independent constant, i.e., it is the power of the integral
\begin{eqnarray}
\fl &&\int_{\Herm_\odot(k|k)}(\Sdet V)^{N}e^{-\Str V}d[V]\\
\fl&=&\int_{\Herm_\odot(k|k)} \left(\frac{\det(V_{\B\B}-V_{\B\F}V_{\F\F}^{-1}V_{\F\B})}{\det V_{\F\F}}\right)^Ne^{-\tr V_{\B\B}+\tr V_{\F\F}}d[V]\nonumber\\
\fl&=&\int_{\Herm_+(k)} (\det V_{\B\B})^{N-k}e^{-\tr V_{\B\B}}d[V_{\B\B}]\int_{\U(k)} (\det V_{\F\F})^{-N-k}e^{\tr V_{\F\F}}d[V_{\F\F}]\nonumber\\
\fl&&\times\int_{\mathbb{C}^{(k|0)\times(0|k)}}\det\left(\id_{k}-V_{\B\F}V_{\F\B}\right)^Nd[V_{\B\F},V_{\F\B}]\nonumber.
\end{eqnarray}
The latter equality can be found by substituting $V_{\B\F}\to V_{\B\B}V_{\B\F}V_{\F\F}$.
The first two integrals are given by the Selberg integrals~\cite{Mehtabook,Peterbook}
\begin{eqnarray}
\fl \int_{\Herm_+(k)} (\det V_{\B\B})^{N-k}e^{-\tr V_{\B\B}}d[V_{\B\B}]&=&\frac{1}{k!}\left(\prod_{j=0}^{k-1}\frac{\pi^j}{j!}\right) \int_{\mathbb{R}_+^k}\det( x)^{N-k} e^{-\tr x}\Delta_k^2(x)d[x]\nonumber\\
\fl&=&\prod_{j=0}^{k-1}\pi^j(N-j-1)!
\end{eqnarray}
and
\begin{eqnarray}
\fl \int_{\U(k)} (\det V_{\F\F})^{-N-k}e^{\tr V_{\F\F}}d[V_{\F\F}]&=&\frac{1}{k!}\left(\prod_{j=0}^{k-1}\frac{\pi^j}{j!}\right) \int\limits_{[0,2\pi]^k}\hspace*{-0.3cm}\det( e^{i\varphi})^{-N-k} e^{\tr e^{i\varphi}}\Delta_k^2(e^{i\varphi})d[e^{i\varphi}]\nonumber\\
\fl&=&(2\pi i)^k\prod_{j=0}^{k-1}\frac{\pi^j}{(N+j)!},
\end{eqnarray}
where we have first diagonalised the matrices and then integrated over their eigenvalues. To compute the remaining integral over the Grassmann variables, we employ that the Gaussian of Grassmann variables is equal to
\begin{eqnarray}
\fl\int_{\mathbb{C}^{(k|0)\times(0|k)}}\exp\left(-\tr V_{\B\F}V_{\F\B}\right)d[V_{\B\F},V_{\F\B}]&=&\prod_{a,b=1}^k\int (-V_{\B\F,ab}V_{\F\B,ba})dV_{\B\F,ab}dV_{\F\B,ba}\nonumber\\
\fl&=&1.
\end{eqnarray}
Additionally we exploit the superbosonisation formula~\cite{Sommers2007,LSZ2008,KSG2009} which tells us how to replace the product $V_{\B\F}V_{\F\B}$ by a Haar distributed unitary matrix $U\in\U(k)$ at the cost of an additional factor of $\det U^{-k}$. In particular we compute
\begin{eqnarray}
\fl &&\int_{\mathbb{C}^{(k|0)\times(0|k)}}\det\left(\id_{k}-V_{\B\F}V_{\F\B}\right)^Nd[V_{\B\F},V_{\F\B}]\nonumber\\
&=&\frac{\int_{\mathbb{C}^{(k|0)\times(0|k)}}\det\left(\id_{k}-V_{\B\F}V_{\F\B}\right)^Nd[V_{\B\F},V_{\F\B}]}{\int_{\mathbb{C}^{(k|0)\times(0|k)}}\exp\left(-\tr V_{\B\F}V_{\F\B}\right)d[V_{\B\F},V_{\F\B}]}\nonumber\\
\fl&=&\frac{\int_{\U(k)}\det\left(\id_{k}-U\right)^N\det U^{-k}d\mu(U)}{\int_{\U(k)}\exp\left(-\tr U\right)\det U^{-k}d\mu(U)}\nonumber\\
\fl&=&\frac{\int_{[0,2\pi]^k}\det\left(\id_{k}-e^{i\varphi}\right)^N\det( e^{i\varphi})^{-2k}\Delta^2_k(e^{i\varphi})d[e^{i\varphi}]}{\int_{[0,2\pi]^k}\exp\left(-\tr e^{i\varphi}\right)\det( e^{i\varphi})^{-2k}\Delta^2_k(e^{i\varphi})d[e^{i\varphi}]}\nonumber\\
\fl&=&\prod_{j=0}^{k-1}\frac{(N+j)!}{(N-j-1)!}.
\end{eqnarray}
The introduction of the denominator guarantees that we do not get any additional constants from the superbosonisation formula and the diagonalisation of the unitary matrix $U$ as both steps are independent of the integrand. Therefore, the result~\eref{partition.final} simplifies to
\begin{eqnarray}\label{partition.final.b}
\fl  Z_{Y_L^{(M)}}^{(k,N)}(\kappa)&=&\int_{\Herm_\odot^{ML}(k|k)}\Sdet(V_L^{(M)}-\kappa)^{-N}\prod_{m=1}^M\prod_{l=1}^Le^{-\Str V_{l,m}}\frac{d[V_{l,m}]}{(2i)^k\pi^{k^2}}.
\end{eqnarray}

\subsection{Microscopic Tail Statistics}\label{sec:asymptotic}

The tail asymptotic results from combining expression~\eref{partition.final.b} for the partition function~\eref{partition.def} with the scaling shown in~\eref{Meijer-G.micro} for the largest eigenvalues inside the tail. This implies the scaling
\begin{equation}\label{tail.scale}
\kappa=N(c_M\widehat{\lambda})^{-M-1}
\end{equation}
with $c_M$ as in~\eref{unfold.const} and the supermatrix $\widehat{\lambda}$ being fixed. Plugging this into~\eref{partition.final.b}, we can readily carry out the large $N$ limit and find
\begin{eqnarray}\label{partition.asymp.tail}
\fl  &&\lim_{N\to\infty}Z_{Y_L^{(M)}}^{(k,N)}\left(N(c_M\widehat{\lambda})^{-M-1}\right)\\
\fl&=&\lim_{N\to\infty}\int_{\Herm_\odot^{ML}(k|k)}\Sdet\left(\id_{k|k}-\frac{(c_M\widehat{\lambda})^{M+1}}{N}V_L^{(M)}\right)^{-N}\prod_{m=1}^M\prod_{l=1}^Le^{-\Str V_{l,m}}\frac{d[V_{l,m}]}{(2i)^k\pi^{k^2}}\nonumber\\
\fl&=&\int_{\Herm_\odot^{ML}(k|k)}\exp\left[\Str(c_M\widehat{\lambda})^{M+1}V_L^{(M)}\right]\prod_{m=1}^M\prod_{l=1}^Le^{-\Str V_{l,m}}\frac{d[V_{l,m}]}{(2i)^k\pi^{k^2}}\nonumber\\
\fl&=&\left[\int_{\Herm_\odot^{M}(k|k)}\exp\left[\Str\left(\frac{c_M\widehat{\lambda}}{L}\right)^{M+1}V_{1,1}^{-1}\cdots V_{1,M}^{-1}\right]\prod_{m=1}^Me^{-\Str V_{1,m}}\frac{d[V_{1,m}]}{(2i)^k\pi^{k^2}}\right]^L.\nonumber
\end{eqnarray}
Certainly, this limit works only when $\widehat{\lambda}_{\B\B}^{M+1}+[\widehat{\lambda}_{\B\B}^{M+1}]^\dagger<0$ is negative definite. This is usually not the case since the imaginary part of $\widehat{\lambda}$ should eventually be set to $0$. Therefore we deform the contours by a slight rotation $V_{1,j}\to e^{-i\pi /(M+1)}V_{1,j}$ so that for a positive definite Hermitian $\widehat{\lambda}$ we obtain the convergent result
\begin{eqnarray}\label{partition.asymp.tail.b}
\fl  \lim_{N\to\infty}Z_{Y_L^{(M)}}^{(k,N)}\left(N(c_M\widehat{\lambda})^{-M-1}\right)&=&\biggl[\int_{\Herm_\odot^{M}(k|k)}\prod_{m=1}^M\frac{d[V_{1,m}]}{(2i)^k\pi^{k^2}}\\
\fl&&\hspace*{-3cm}\times\exp\left[-e^{-i\pi/(M+1)}\Str\left(\left[\frac{c_M\widehat{\lambda}}{L}\right]^{M+1}V_{1,1}^{-1}\cdots V_{1,M}^{-1}+\sum_{m=1}^MV_{1,m}\right)\right]\biggl]^L\nonumber\\
\fl&=& \left[\lim_{N\to\infty}Z_{Y_1^{(M)}}^{(k,N)}\left(N\left[\frac{L}{c_M\widehat{\lambda}}\right]^{M+1}\right)\right]^L.\nonumber
\end{eqnarray}
In the last line, we recognise our claim in Sec.~\ref{sec:observe} that the tail eigenvalue statistics agree with those of a direct sum, which is namely
\begin{eqnarray}
\fl  \left\langle\prod_{j=1}^k\frac{\det(\bigoplus_{l=1}^L(X_l^{(M)})^\dagger X_l^{(M)}-\kappa_{\F,j}\otimes\id_L)}{\det(\bigoplus_{l=1}^L(X_l^{(M)})^\dagger X_l^{(M)}-\kappa_{\B,j}\otimes\id_L)}\right\rangle&=&\left[\left\langle\prod_{j=1}^k\frac{\det((X_1^{(M)})^\dagger X_1^{(M)}-\kappa_{\F,j})}{\det((X_1^{(M)})^\dagger X_1^{(M)}-\kappa_{\B,j})}\right\rangle\right]^L\nonumber\\
\fl&=& \left[Z_{Y_1^{(M)}}^{(k,N)}(\kappa)\right]^L.\label{partition.direct.tail}
\end{eqnarray}
For $M=L=1$, this result agrees with the Bessel kernel result~\cite{Kieburg2015} which has been an important result in the study of Quantum Chromodynamics~\cite{Verbaarschot2004}.

The result~\eref{partition.asymp.tail.b} is the analytical corroboration which we previously mentioned despite that we considered here a particular kind of ensemble where we could carry out the computation. However, we are rather sure that this is the generic behaviour of heavy-tailed statistics. The eigenvalues seem to be too diluted to show their level repulsion which is reflected in the Vandermonde determinant of their joint probability densities, so that effectively they behave as if this level repulsion never existed.

\subsection{Critical Regime of the Spectral Statistics}\label{sec:critical}

Now we investigate how far into the bulk the limiting statistics to a direct sum of random matrices carries over and what the critical regime is where we eventually enter stable statistics such as the sine kernel in the bulk. For this aim we consider a general scaling
\begin{equation}\label{general.scale}
\kappa=N^\gamma \kappa_0\id_{k|k}+N^\delta\widetilde{\kappa}
\end{equation}
with $\delta<\gamma<1$. The point $\kappa_0>0$ is where we zoom into the spectrum and hence the condition $\delta<\gamma$ where $\delta$ has to be determined so that the spectral fluctuations on the scale of the local mean level spacing are resolved. For $\gamma=1$ this scale has been $\delta=\gamma$ as we have seen and for $\gamma>1$ we look into the tail of the largest eigenvalue only implying that we do not see anything of much interest.

Choosing the slightly rotated version of the supermatrices $V_{l,m}$, cf., Eq.~\eref{partition.asymp.tail.b}, we rescale the supermatrices by $N^{(1-\gamma)/(M+1)}$
\begin{eqnarray}
\fl  &&Z_{Y_L^{(M)}}^{(k,N)}\left(N^\gamma \kappa_0\id_{k|k}+N^\delta\widetilde{\kappa}\right)\\
\fl&=&\int_{\Herm_\odot^{ML}(k|k)}\Sdet\left(\id_{k|k}+e^{-i\pi/(M+1)}N^{-(M+\gamma)/(M+1)}( \kappa_0\id_{k|k}+N^{\delta-\gamma}\widetilde{\kappa})^{-1}V_L^{(M)}\right)^{-N}\nonumber\\
\fl&&\times\prod_{m=1}^M\prod_{l=1}^L\exp\left[-e^{-i\pi/(M+1)}N^{(1-\gamma)/(M+1)}\Str V_{l,m}\right]\frac{d[V_{l,m}]}{(2i)^k\pi^{k^2}}.\nonumber
\end{eqnarray}
When Taylor expanding the logarithm of the superdeterminant we notice that only the linear term survives in the large $N$ limit since $1-j(M+\gamma)/(M+1)<0$ for all $j\geq2$ and any $M\geq 1$ and $\gamma>1-M$, so that
\begin{eqnarray}\label{direct.sum.limit}
\fl  &&Z_{Y_L^{(M)}}^{(k,N)}\left(N^\gamma \kappa_0\id_{k|k}+N^\delta\widetilde{\kappa}\right)\\
\fl&\overset{N\gg1}{\approx}&\int_{\Herm_\odot^{ML}(k|k)}\exp\left[-e^{-i\pi/(M+1)}N^{(1-\gamma)/(M+1)}\Str( \kappa_0\id_{k|k}+N^{\delta-\gamma}\widetilde{\kappa})^{-1}V_L^{(M)}\right]\nonumber\\
\fl&&\times\prod_{m=1}^M\prod_{l=1}^L\exp\left[-e^{-i\pi/(M+1)}N^{(1-\gamma)/(M+1)}\Str V_{l,m}\right]\frac{d[V_{l,m}]}{(2i)^k\pi^{k^2}}\nonumber\\
\fl&\approx&\left[Z_{Y_1^{(M)}}^{(k,N)}\left(L^{M+1}(N^\gamma \kappa_0\id_{k|k}+N^\delta\widetilde{\kappa})\right)\right]^L.\nonumber
\end{eqnarray}
In this expression, we can read off the local scale given by the exponent $\delta=[(M+2)\gamma-1]/(M+1)$ as then the Taylor expansion in $\widetilde{\kappa}$ terminates with the linear term  in the asymptotic limit $N\to\infty$.

Equation~\eref{direct.sum.limit} already shows that even some part of the bulk statistics close to the tail still follows the statistics of a direct sum. A more detailed saddle point analysis would show that we get the direct sum of $L$ independent sine-kernel statistics. The condition of the scaling exponent $\gamma$ which has to be satisfied for these kind of statistics is $\gamma>1-M$. Hence, for $\gamma\leq 1-M$ we have to go to higher order expansions of the logarithm of the superdeterminant. Those terms couple the supermatrices.

For $\gamma<1-M$, those higher order terms are also large so that we need to carry out an additional saddle point expansion where all $V_{l,m}$ become equal. Thence we would find the statistics of a single sine-kernel, see~\cite{Zirnbauer1996,Verbaarschot2004} for the supersymmetric integral expression of these statistics. One needs to be careful, as Rothstein vectorfields~\cite{Rothstein1987} will occur in the saddle point expansion as they account for all Efetov-Wegner boundary terms~\cite{Wegner1983,Efetov1983} that correspond to the diagonalisation of a supermatrix. All those terms have been explicitly computed for diagonalising Hermitian supermatrices in~\cite{Kieburg2011}.

For $\gamma=1-M$ and, hence, $\delta=(1-M-M^2)/(M+1)$, the quadratic term of the Taylor expansion is of order $1$ so that all supermatrices are coupled, 
\begin{eqnarray}\label{critical.limit}
\fl  &&Z_{Y_L^{(M)}}^{(k,N)}\left(N^{1-M} \kappa_0\id_{k|k}+N^{(1-M-M^2)/(M+1)}\widetilde{\kappa}\right)\\
\fl&\overset{N\gg1}{\approx}&\int_{\Herm_\odot^{ML}(k|k)}\exp\left[-e^{-i\pi/(M+1)}N^{M/(M+1)}\Str\left( \kappa_0^{-1}V_L^{(M)}+\sum_{m=1}^M\sum_{l=1}^LV_{l,m}\right)\right]\nonumber\\
\fl&&\times\exp\left[\frac{e^{-i\pi/(M+1)}}{\kappa_0^{2}}\Str \widetilde{\kappa}V_L^{(M)}+\frac{e^{-2i\pi/(M+1)}}{2\kappa_0^{2}}\Str(V_L^{(M)})^2\right]\prod_{m=1}^M\prod_{l=1}^L\frac{d[V_{l,m}]}{(2i)^k\pi^{k^2}}.\nonumber
\end{eqnarray}
This expression can be decoupled by a Hubbard-Stratonovich transformation~\cite{Hubbard1959,Stratonovich1957}
\begin{eqnarray}\label{HS.trafo}
  &&\exp\left[\frac{e^{-2i\pi/(M+1)}}{2\kappa_0^{2}}\Str(V_L^{(M)})^2\right]\\
&=&\frac{\int_{\Herm(k|k)}\exp[-\frac{1}{2}\Str \Xi^2+\frac{e^{-i\pi/(M+1)}}{\kappa_0}\Str\Xi V_L^{(M)}]d[\Xi]}{\int_{\Herm(k|k)}\exp[-\frac{1}{2}\Str \Xi^2]d[\Xi]},\nonumber
\end{eqnarray}
with $\Xi$ a supermatrix whose Boson-Boson block is an arbitrary Hermitian matrix $\Xi_{\B\B}=\Xi_{\B\B}^\dagger\in\Herm(k)$ and the Fermion-Fermion block is an arbitrary anti-Hermitian matrix $\Xi_{\F\F}=-\Xi_{\F\F}^\dagger\in i\Herm(k)$. The Boson-Fermion and Fermion-Boson blocks again contain independent Grassmann variables and the measure $d[\Xi]$ is the product of all differentials of the supermatrix entries. The partition function can then be approximated anew by the partition function of a direct sum of identical random matrices which however are now coupled by the supermatrix $\Xi$, i.e.,
\begin{eqnarray}\label{critical.limit.SUSY.coupling}
 &&Z_{Y_L^{(M)}}^{(k,N)}\left(N^{1-M} \kappa_0\id_{k|k}+N^{(1-M-M^2)/(M+1)}\widetilde{\kappa}\right)\\
&\overset{N\gg1}{\approx}&\frac{\int_{\Herm(k|k)}[Z_{Y_1^{(M)}}^{(k,N)}\left(K(\Xi)\right)]^L\exp[-\frac{1}{2}\Str \Xi^2]\ d[\Xi]}{\int_{\Herm(k|k)}\exp[-\frac{1}{2}\Str \Xi^2]d[\Xi]}\nonumber
\end{eqnarray}
with
\begin{equation}
K(\Xi)=L^{M+1}[N^{1-M} \kappa_0\id_{k|k}+N^{(1-M-M^2)/(M+1)}(\widetilde{\kappa}+\kappa_0\Xi)].
\end{equation}
This result resembles those in~\cite{KVZ2013} where the transition between independent diagonal blocks of random matrices between a full matrix have been considered. This underlines that our understanding of a transition between a direct sum to a full matrix without block structures in the tail is ostensibly correct.

One last comment on the critical scale of the eigenvalues $\lambda\propto N^{1-M}$ of the  random matrix $Y_L^{(M)}$. It is only slightly larger than the scale $N^{-M}$ of the macroscopic level density~\eref{MP.M-law} which is freely stable. Although we are already then deep in the tail we are far away from the scale of the largest eigenvalue which scales like $N$ for the chosen reference scale of the product of inverse Ginibre ensembles, cf., Eq.~\eref{Inv-Gin}. We believe that the relative scales should hold for other ensembles too and that in particular the ratio of the scale between the largest eigenvalues and the critical scale should follow the law ${\rm scale}_{\rm largest\ eigenvalue}/{\rm scale}_{\rm critical}=N^{1/(2\alpha)}$ where $\alpha$ is the stability exponent. The Taylor expansion should follow the same mechanism however the probability density $P(H)$ and thus the corresponding superfunction $Q(U)$ may vary.

\section{Stable Ensembles and Poisson  Statistics in the Tail}\label{sec:poisson}

The last thing we would like to address in the present article is the limit $L\to\infty$. The multivariate central limit theorem~\cite{Rvaceva1954}, for unitarily invariant random matrix ensembles see~\cite{ZK2021}, tells us that if the limit exists it should converge to one of the stable random matrix ensembles and this already occurs at finite matrix dimension $N$. Specifically this means when the Hermitian matrix $H\in\Herm(N)$ is a strictly stable random matrix associated to the stability exponent $\alpha$ and we draw two copies $H_1$ and $H_2$ of  $H$, then the sum $(H_1+H_2)/2^{1/\alpha}$ is also a copy of $H$, implying that it exhibits the very same statistics, including but not limited to eigenvalues, eigenvectors, and matrix entries (also for finite $N$). Hence this behaviour should carry over to the large $N$ limit. The only question is whether the two limits $L\to\infty$ and $N\to\infty$ commute. Our numerical and analytical simulations suggest that this might be true for some cases of the microscopic and macroscopic spectral scales depending on the averaged position of the largest eigenvalues in the tail.

Additionally, mesoscopic spectral scales might arise which are reminiscent to the order of the two limits. What underlines the latter point is the fact that the Bessel result for the level density~\eref{hard-edge.micro} of the case $M=1$ can be found for all $L$, see left plots in Fig.~\ref{fig:rhotailind-alp05-bet1}. One needs only to cluster the eigenvalues in $L$ consecutive couples.

The limit $L\to\infty$ for the model considered in Sec.~\ref{sec:analytical} and in particular for the result~\eref{partition.direct.tail},  is carried out in subsection~\ref{sec:limit}, while we consider a random matrix ensemble that is already stable at finite $N$ in subsection~\ref{sec:stable}.

\subsection{Limiting Statistics for the Model of Sec.~\ref{sec:analytical}}\label{sec:limit}

What does the discussion above on the limiting stable distributions imply for the tail statistics? For that reason, we  start from the knowledge that the microscopic tail statistics follows approximately the one of a direct sum of random matrices, i.e., we start from~\eref{partition.asymp.tail.b}. The spectral variables are scaled as follows
\begin{equation}
\kappa=N\left[L\left(\lambda_0\id_{k|k}+\frac{\widetilde{\lambda}}{\lambda_0^M(M+1)L}\right)\right]^{-M-1},
\end{equation}
where $\lambda_0>0$ is the base point where we zoom into the spectrum and $\widetilde{\lambda}$ measures the spectral fluctuations. Plugging this into~\eref{partition.asymp.tail.b} and expanding for large $L$, we obtain 
\begin{eqnarray}
\fl&&  \lim_{L\to\infty}\lim_{N\to\infty}Z_{Y_L^{(M)}}^{(k,N)}\left(\kappa\right)\nonumber\\
\fl&=&\lim_{L\to\infty}\biggl(\int\limits_{\Herm_\odot^{M}(k|k)}\prod_{m=1}^M\frac{d[V_{1,m}]}{(2i)^k\pi^{k^2}}\left(1-\frac{1}{L}e^{-i\frac{\pi}{M+1}}\Str \widetilde{\lambda}V_{1,1}^{-1}\cdots V_{1,M}^{-1}\right)\nonumber\\
\fl&&\times\exp\left[-e^{-i\pi/(M+1)}\Str\left(\lambda_0^{M+1}V_{1,1}^{-1}\cdots V_{1,M}^{-1}+\sum_{m=1}^MV_{1,m}\right)\right]\biggl)^L\nonumber\\
\fl&=&\exp\biggl(-\int\limits_{\Herm_\odot^{M}(k|k)}\prod_{m=1}^M\frac{d[V_{1,m}]}{(2i)^k\pi^{k^2}}e^{-i\frac{\pi}{M+1}}\Str \widetilde{\lambda}V_{1,1}^{-1}\cdots V_{1,M}^{-1}\nonumber\\
\fl&&\times\exp\left[-e^{-i\pi/(M+1)}\Str\left(\lambda_0^{M+1}V_{1,1}^{-1}\cdots V_{1,M}^{-1}+\sum_{m=1}^MV_{1,m}\right)\right]\biggl).\label{partition.largeL.tail.a}
\end{eqnarray}
For the second equality we have exploited the fact that the integrand is normalised for any $\kappa\propto\id_{k|k}$ because then all determinants in~\eref{partition.def} cancel. The average in the exponent can be simplified due to the supergroup invariance of the integrand that is only broken by $\Str \widetilde{\lambda}V_{1,1}^{-1}\cdots V_{1,M}^{-1}$ so that eventually we have
\begin{eqnarray}\label{partition.largeL.tail.b}
 \lim_{L\to\infty}\lim_{N\to\infty}Z_{Y_L^{(M)}}^{(k,N)}\left(\kappa\right)&=&\exp[C\Str\widetilde{\lambda}]
\end{eqnarray}
with a constant $C$ that depends only on $k$, $M$ and $\lambda_0$.

Let us compare~\eref{partition.largeL.tail.b} with the partition function of a Poisson distributed spectrum, meaning the random matrix $H=\diag(E_1,\ldots,E_N)$ is diagonal and each eigenvalue $E_j$ is independently and identically distributed by $F(E)$. We consider the average
\begin{equation}\label{Poisson.partition}
\fl Z_{\rm Poisson}^{(k,N)}(\kappa)=\left\langle\Sdet(H\otimes\id_{k|k}-\id_N\otimes\kappa)^{-1}\right\rangle=\left[\left\langle\Sdet(E\id_{k|k}-\kappa)^{-1}\right\rangle\right]^N,
\end{equation}
which is the counterpart of~\eref{partition.sdet}. The average on the right hand side is over a single eigenvalue only.

The local spectral fluctuations of a Poisson ensemble happen on the scale $1/N$ when the distribution $F(E)$ is $N$ independent. Therefore, we choose the scaling $\kappa=\lambda_0\id_{k|k}+\widetilde{\lambda}/N$ with $\lambda_0$ the base point with a tiny imaginary increment for the regularization then the limit $N\to\infty$ of~\eref{Poisson.partition} leads to
\begin{equation}\label{Poisson.partition.limit}
 \lim_{N\to\infty}Z_{\rm Poisson}^{(k,N)}(\kappa)=\exp\left[-\int \frac{F(E)dE}{E-\lambda_0}\Str\widetilde{\lambda}\right].
\end{equation}
Comparison with the result~\eref{partition.largeL.tail.b} underlines our point that in the large $L$ limit we indeed find the Poisson statistics.

Since the critical scaling $\kappa\propto N^{1-M}$ where the transition to the sine-kernel statistics happens, is independent of $L$ we expect that it is the same critical scale where the Poisson statistics should turn over into the sine-kernel as well. This certainly deserves more investigation, yet we skip it here as it exceeds the scope of the present work.

\subsection{Tail Statistics for a Stable Random Matrix Model}\label{sec:stable}

To understand better whether the Poisson statistic holds true generally in the heavy tail as we know it does for heavy-tailed Wigner ensembles~\cite{Soshnikov2004,BBP2007,ABAP2009,TBT2016}, we have also numerically checked whether an already stable heavy-tailed unitarily invariant random matrix exhibits the Poisson statistics in its tail. Therefore we would like to point out that such an ensemble can be readily constructed with the help of the GUE. Its probability density is
\begin{equation}\label{GUE.def}
P_{\rm GUE}(H;\sigma)=\frac{\exp[-\tr H^2/(2\sigma^2)]}{2^{N/2}(\pi\sigma^2)^{N^2/2}},\qquad H\in\Herm(N),
\end{equation}
with an arbitrary standard deviation $\sigma>0$. This standard deviation is now drawn from a stable totally asymmetric univariate density with a stability exponent $\widetilde{\alpha}\in]0,1[$ and asymmetry parameter $\widetilde{\beta}=1$. The distribution of such a density is given by the Fourier transform~\cite{Levy}
\begin{equation}\label{stable.uni}
 \widehat{p}_{\widetilde{\alpha}}(x)=\int_{-\infty}^\infty \exp\left[-ix\omega-\frac{(-i\omega)^{\widetilde{\alpha}}}{\cos(\pi\widetilde{\alpha}/2)}\right]\frac{d\omega}{2\pi}.
\end{equation}
The condition $\widetilde{\alpha}<1$ is important because only then the support is restricted on the positive real line $\mathbb{R}_+$.

When determining the variance $\sigma^2$ from $p_{\widetilde{\alpha}}(x)$, we obtain a symmetric unitarily invariant random matrix which is stable with stability exponent $\alpha=2\widetilde{\alpha}$. In particular we consider the random matrix distribution
\begin{equation}\label{GUE.stable}
P_{\alpha}(H)=\int_0^\infty\frac{\exp[-\tr H^2/(2x)]}{2^{N/2}(\pi x)^{N^2/2}} \widehat{p}_{\alpha/2}(x)dx.
\end{equation}
A simple computation readily proves its stability
\begin{eqnarray}
\fl 2^{N^2/\alpha}P_{\alpha}\ast P_{\alpha}(2^{1/\alpha}H)&=&2^{N^2/\alpha}\int_{\Herm(N)}P_{\alpha}(H') P_{\alpha}(2^{1/\alpha}H-H')d H'\nonumber\\
\fl&=&\int_0^\infty dx_1\int_0^\infty dx_2 \frac{\exp\left[-\frac{\tr H^2}{2^{1-2/\alpha}[x_1+x_2]}\right]}{2^{N/2}(2^{-2/\alpha}\pi [x_1+x_2])^{N^2/2}}\widehat{p}_{\frac{\alpha}{2}}(x_1)\widehat{p}_{\frac{\alpha}{2}}(x_2)\nonumber\\
\fl&=&P_{\alpha}(H).
\end{eqnarray}
The factor $2^{N^2/\alpha}$ is the Jacobian due to the rescaling $H\to 2^{1/\alpha}H$. Additionally the second equality results from the convolution rules of two Gaussians and the third equality takes into account the stability of $\widehat{p}_{\alpha/2}(x)$.

The construction~\eref{GUE.stable}, where one averages over the variance has also been studied for similar ensembles in~\cite{BGW2006,BCP2008,AFV2010,AV2008,AMAV2009} though the authors of this work did not aim for stable distributions. 

We have employed the construction above for the Monte Carlo simulations which readily can be numerically generated by noticing that $\sqrt{\sigma} H$ with $\sigma$ and $H$ independently drawn from $ \widehat{p}_{\alpha/2}(x)$ and $P_{\rm GUE}(H;1)$, respectively, leads to the same random matrix. For four different stability exponents $\alpha=0.5,1,1.5,1.8$, we have numerically simulated the level spacing distribution as well as their level density for the four largest and four smallest eigenvalues. Those are drawn in Fig.~\ref{fig:Spacing-distributions-stable}. Since the ensemble is symmetric about the origin the ensemble exhibits on both sides heavy-tails. 

The macroscopic level density is an averaged Wigner semicircle
\begin{equation}\label{macro.density.alpha}
\fl\rho_\alpha(\lambda)=\lim_{N\to\infty}\frac{1}{N}\left\langle\tr\delta(\lambda\id_N-N^{-1/2}H)\right\rangle=\int_{\lambda^2/4}^\infty \frac{\sqrt{4x-\lambda^2}}{2\pi x}\widehat{p}_{\alpha/2}(x)dx.
\end{equation}
Evidently, the asymptotic behaviour is $\rho_\alpha(\lambda)\propto|\lambda|^{-1-\alpha}$ for $|\lambda|\gg1$.

\begin{figure}[t!]
\begin{center}
\includegraphics[width=0.8\textwidth]{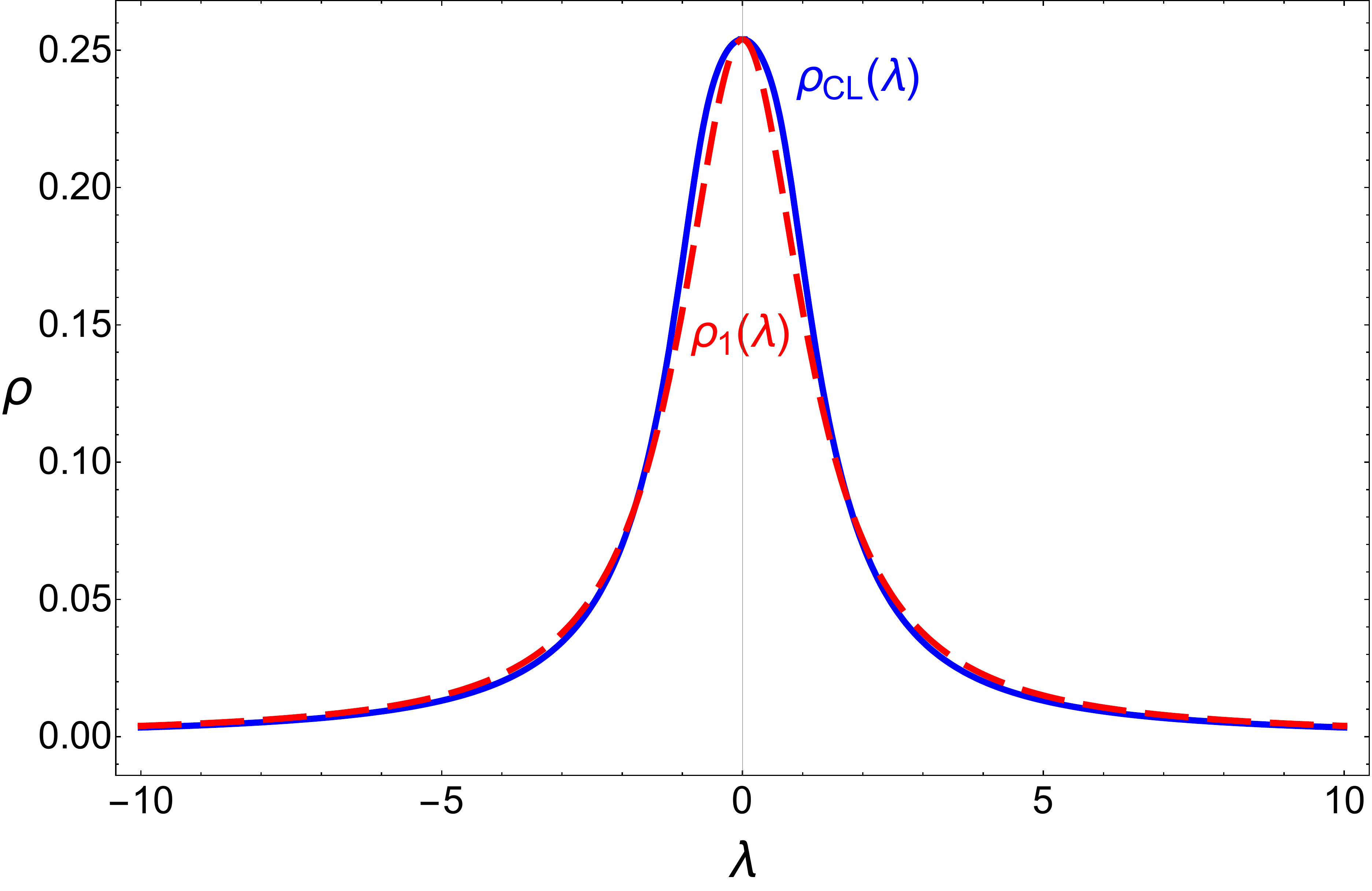}
\caption{The macroscopic level densities $\rho_1(\lambda)$ (blue solid curve, see Eq.~\eref{macro.density.alpha}) and  $\rho_{\rm CL}(\lambda)$ (red dashed curve, see Eq.~\eref{Lorentz}). Despite the corresponding ensembles are stable only those yielding $\rho_{\rm CL}(\lambda)$ can be stable under free convolution.}
\label{fig:free-vs-non-free}
\end{center}
\end{figure}

Despite the considered ensemble being stable, it is not 'freely stable' meaning two copies of the matrix are not free random variables. One can convince oneself fairly easily by considering the case $\alpha=1$. The symmetric distribution which is stable under free convolution is uniquely given (up to a scaling parameter $c>0$) by the Lorentz function~\cite{BPB1999} (also known as the Cauchy or Breit-Wigner distribution)
\begin{equation}\label{Lorentz}
\rho_{\rm CL}(\lambda)=\frac{1}{\pi}\frac{c}{1+c^2\lambda^2}.
\end{equation}
One can then show that it is always the case $\rho_1(\lambda)\neq \rho_{\rm CL}(\lambda)$ regardless of what scaling $c>0$  is chosen, cf., Fig.~\ref{fig:free-vs-non-free} where $c$ is fixed by $\rho_1(0)=c/\pi$.

\begin{figure}[t!]
\begin{center}
\includegraphics[width=\textwidth]{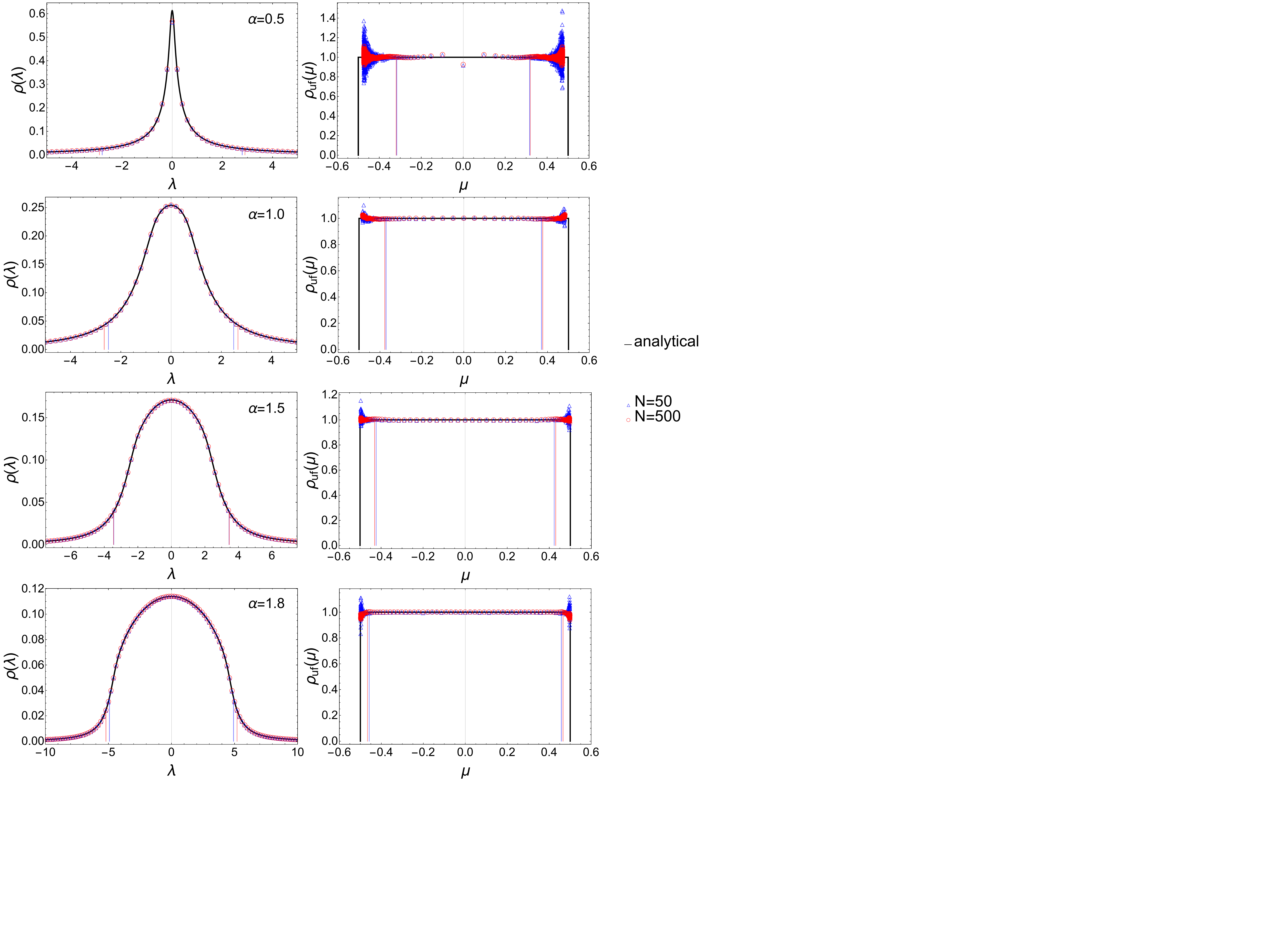}
\caption{The non-unfolded (left plots) and unfolded (right plots) macroscopic level densities for the stable ensemble~\eref{GUE.stable} for the four stability exponents $\alpha=0.5,1.0,1.5,1.8$ and the matrix dimensions $N=50,500$. The analytical curves (solid black curves) are generated via the integral~\eref{macro.density.alpha}. The unfolded eigenvalues $\mu$ are given by~\eref{unfolding}. For each ensemble (coloured symbols) we have created $10^5$ configurations. The scattering in the unfolded densities close to the boundaries $\mu=\pm1/2$ can be explained due to very rare events as those points correspond to the heavy tails. The slight dip at $\mu=0$ for $\alpha=0.5$ can be understood by the very narrow peak which is not sufficiently resolved by the chosen bin. The vertical lines show the mean positions of the largest and smallest eigenvalue in the unfolded variables $\mu$. In the left plots we have mapped these position back to $\lambda$.}
\label{fig:level-density-stable}
\end{center}
\end{figure}

\begin{figure}[t!]
\begin{center}
\includegraphics[width=\textwidth]{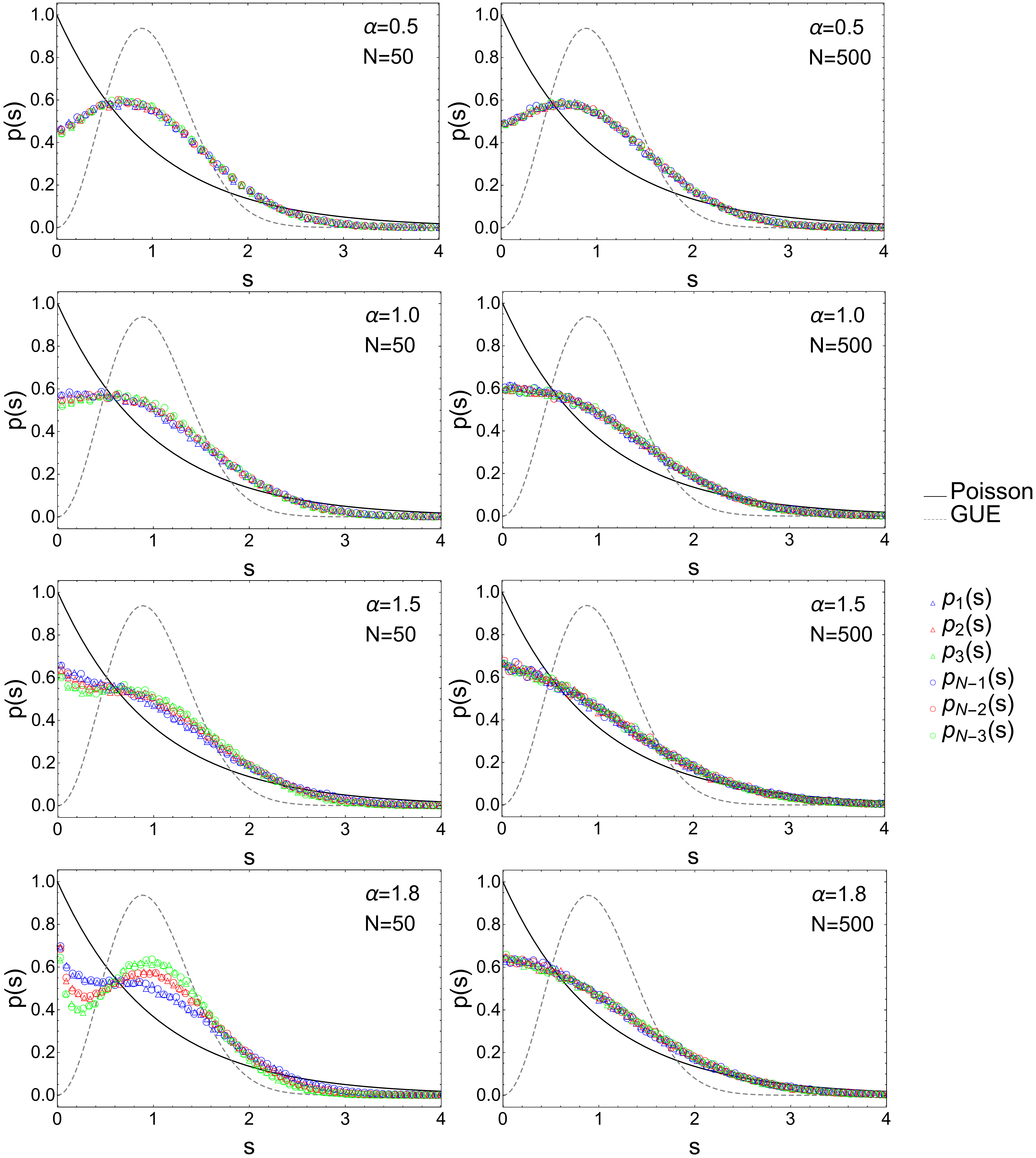}
\caption{The level spacing distributions of the unfolded eigenvalues~\eref{unfolding} for the stable random matrix ensemble~\eref{GUE.stable} for the four stability exponents $\alpha=0.5,1.0,1.5,1.8$ and the matrix dimensions $N=50,500$. The Monte Carlo simulations (coloured symbols) are the same of Fig.~\ref{fig:level-density-stable}. The nomenclature $p_k(s)$ refers to the spacing of the $k$th and $(k+1)$st smallest eigenvalue. Due to symmetry of the ensemble, the two distributions $p_k(s)$ and $p_{N-k}(s)$ must agree as they indeed approximately do (same coloured circles and triangles).  As a comparison we have drawn the Poisson distribution~\eref{Poisson-dist} (solid curve) and the Wigner surmise~\eref{GUE-surmise-dist} (dashed curve).}
\label{fig:Spacing-distributions-stable}
\end{center}
\end{figure}

When computing the (unfolded) largest eigenvalues, we have been very surprised by the fact that the largest eigenvalues are still deep in the bulk of the distribution $\rho_\alpha(\lambda)$ in contrast to the model in Sec.~\ref{sec:analytical}. Thence, the approximation of this distribution by its leading asymptotic behaviour $\rho_\alpha(\lambda)\propto\lambda^{-1-\alpha}$ has not been suitable for the unfolding. Instead, we have unfolded the entire spectrum with the new variables
\begin{equation}\label{unfolding}
\mu(\lambda)=\int_0^\lambda \rho_\alpha(\lambda')d\lambda'.
\end{equation}
This substitution maps the spectrum from the real line $\mathbb{R}$ to the open interval $]-1/2,1/2[$ with a uniform level density. The mean of the largest eigenvalue after this mapping is well-defined and after mapping this back to the original spectrum, they have been indicated as vertical lines in Fig.~\ref{fig:level-density-stable}.

The reason why the largest eigenvalues do not lie at the utmost end of the heavy-tail (for the unfolded variables at $\mu=\pm1/2$) is that the tails are a superposition of almost all eigenvalues instead of (essentially) only the largest one. Therefore the situation is not clear as to whether we should find the Poisson statistics when taking the limit $N\to\infty$. To further understand the problem in more detail we compute the partition function
\begin{eqnarray}\label{stable.partition}
 Z_{\alpha}^{(k,N)}(\kappa)&=&\left\langle\Sdet(H\otimes\id_{k|k}-\id_N\otimes\kappa)^{-1}\right\rangle\\
&=&\int_0^\infty\left\langle\Sdet(H\otimes\id_{k|k}-\id_N\otimes\kappa)^{-1}\right\rangle_{x}\widehat{p}_{\alpha/2}(x)dx\nonumber
\end{eqnarray}
with respect to the considered stable ensemble. The notation $\langle.\rangle_{x}$ denotes the average over the GUE with variance $x$.

The average over the GUE can be cast into a supermatrix integral like in~\cite{KSG2009} with the help of the superbosonisation formula.
In doing so we assume that the Boson-Boson block $\kappa_{\B\B}$ is diagonalised which is always possible when its Jordan normal form is diagonal. Then we define the diagonal matrix $\widehat{S}=({\rm sign}({\rm Im}[\kappa_{\B\B}]),\id_k)$ which comprises all signs of the imaginary parts of $\kappa_{\B\B}$. In this way, the Hermitian numerical part of $(iH\otimes\widehat{S}-i \id_N\otimes\widehat{S}\kappa)_{\B\B}$ is positive definite. This allows us to write the superdeterminant in terms of an average over a Gaussian integral of a rectangular supermatrix $V$ of size $(N|0)\times(k|k)$ as the convergence is guaranteed now,
\begin{equation}
\fl\Sdet(H\otimes\id_{k|k}-\id_N\otimes\kappa)^{-1}=\frac{\int \exp[i \Str V^\dagger V\widehat{S}\kappa-i\tr HV\widehat{S}V^\dagger]d[V]}{\int \exp[-\Str V^\dagger V]d[V]}.
\end{equation}
After averaging over $H$, we arrive at
\begin{equation}
\fl\left\langle\Sdet(H\otimes\id_{k|k}-\id_N\otimes\kappa)^{-1}\right\rangle_{x}=\frac{\int \exp[i \Str V^\dagger V\widehat{S}\kappa-x\Str (\widehat{S}V^\dagger V)^2/2]d[V]}{\int \exp[-\Str V^\dagger V]d[V]}.
\end{equation}
In the last step, we employ the superbosonisation formula~\cite{Sommers2007,LSZ2008,KSG2009} and replace $V^\dagger V$ by $\gamma U$ with $\gamma>0$ a scaling that needs to be adjusted and the supermatrix $U\in\Herm_\odot(k|k)$ which is realized from the very same set as in the supersymmetric projection formula~\eref{SUSY.projection}. Thus we eventually arrive at
\begin{eqnarray}\label{stable.partition.b}
Z_{\alpha}^{(k,N)}(\kappa)&=&\int_0^\infty dx\int_{\Herm_\odot(k|k)} \frac{d[U]}{(2i)^k\pi^{k^2}}\widehat{p}_{\alpha/2}(x)\Sdet U^N\\
&&\times\exp\left[-\frac{x\gamma^2}{2}\Str (\widehat{S}U)^2+i \gamma\Str U\widehat{S}\kappa\right].\nonumber
\end{eqnarray}

The scale of the macroscopic level density is obtained for $\kappa\propto\sqrt{N}\kappa_0\in\mathbb{R}$. Then we choose $\gamma=\sqrt{N}$ and perform the saddle point analysis for $N\gg1$. The two corresponding saddle point solutions for the eigenvalues of $U\widehat{S}$ are $z_\pm=(i\kappa_0\pm\sqrt{4x-\kappa_0^2})/(2x)$. Here we have to split the discussion into cases depending on whether $4x$ is larger or smaller than $\kappa_0^2$.

If $4x>\kappa_0^2$, then the real parts of the eigenvalues in the Boson-Boson block of $U\widehat{S}$ must have the signature $\widehat{S}_{\B\B}$ as otherwise the maximum of the integrand along the contour is never acquired at the saddle point. Moreover the eigenvalues of the Fermion-Fermion block need to be the same eigenvalues since only then the Berezinian of the diagonalisation of $U\widehat{S}$ (Jacobian in superspace, see~\cite{Berezinbook}) is of order one. Other solutions will yield higher orders in $1/N$. To summarise the saddle point manifold is given by
\begin{equation}\label{solution.xgreaterkappa}
U\widehat{S}=\frac{i\kappa_0}{2x}\id_{k|k}+\frac{\sqrt{4x-\kappa_0^2}}{2x}\widetilde{U}\diag(\widehat{S}_{\B\B},\widehat{S}_{\B\B})\widetilde{U}^{-1}
\end{equation}
where $\widetilde{U}\in\U(k_+,k_-|k_+,k_-)/[\U(k_+|k_+)\times\U(k_-|k_-)]$ is a Haar distributed unitary supermatrix with $k_+$ and $k_-$ the number of plus and minus signs in $\widehat{S}_{\B\B}$. We would like to point out that the change of coordinates for~\eref{solution.xgreaterkappa} involves a Rothstein vectorfield~\cite{Rothstein1987} which we denote by $\mathcal{Y}_{\widetilde{U}}$. It is however $N$-independent because it only corresponds to the substitution and not the integrand as we know it for Jacobians.

If $4x\leq\kappa_0^2$, the solutions become entirely imaginary though only one of them is a maximum of the integrand for the integration variables. The second derivative of the exponential term is at the two saddle points
\begin{equation}
\partial^2_z\left.\left(-\frac{xz^2}{2}+i\kappa_0 z+{\rm ln}(z)\right)\right|_{z=z_\pm}=-x+\frac{4x^2}{(\kappa_0\pm\sqrt{\kappa_0^2-4x})^2}.
\end{equation}
Combining this with the fact that the bosonic eigenvalues run through these points along the imaginary line and the fermionic ones parallel to the real line, both amount to an additional minus sign in the second term of the Taylor expansion about the saddle point in~\eref{stable.partition.b}. Therefore only
\begin{equation}\label{solution.xlesskappa}
U\widehat{S}=i\frac{\kappa_0-\sqrt{\kappa_0^2-4x}}{2x}\id_{k|k}
\end{equation}
can be a maximum along the contours. Here, we would like to underline that no Rothstein vectorfield is needed as we do not need to diagonalise the supermatrix to reach this saddle point in contrast to the case $4x>\kappa_0^2$.

The spectral fluctuations can be obtained by setting $\kappa=\sqrt{N}\kappa_0\id_{k|k}+\widetilde{\kappa}/\sqrt{N}$. We expand
\begin{equation}
U\widehat{S}=\frac{i\kappa_0}{2x}\id_{k|k}+\frac{\sqrt{4x-\kappa_0^2}}{2x}\widetilde{U}\diag(\widehat{S}_{\B\B},\widehat{S}_{\B\B})\widetilde{U}^{-1}+\frac{1}{\sqrt{N}}\widetilde{U}\delta Q\widetilde{U}^{-1}
\end{equation}
for $4x>\kappa_0^2$ and
\begin{equation}
U\widehat{S}=i\frac{\kappa_0-\sqrt{\kappa_0^2-4x}}{2x}\id_{k|k}+\frac{1}{\sqrt{N}}\delta Q
\end{equation}
for $4x<\kappa_0^2$ up to second order in the massive modes $\delta Q$. The supermatrix $\delta Q$ describes essentially the superspaces $\Herm(k_+|k_+)\times\Herm(k_-|k_-)$ and $\Herm(k|k)$, respectively for the two situations. Their integrations yield $1$ due to the normalisation and we eventually obtain
\begin{eqnarray}\label{stable.partition.c}
\fl\lim_{N\to\infty} Z_{\alpha}^{(k,N)}(\kappa)&=&\int_0^{\kappa_0^2/4} dx\widehat{p}_{\alpha/2}(x)\exp\left[-\frac{\kappa_0-\sqrt{\kappa_0^2-4x}}{2x}\Str \widetilde{\kappa}\right]\\
\fl&&+\int_{\kappa_0^2/4}^\infty dx\widehat{p}_{\alpha/2}(x)\exp\left[-\frac{\kappa_0}{2x}\Str \widetilde{\kappa}\right]\int\limits_{\frac{\U(k_+,k_-|k_+,k_-)}{\U(k_+|k_+)\times\U(k_-|k_-)}} \exp[\mathcal{Y}_{\widetilde{U}}]\nonumber\\
\fl&&\times\exp\left[i\frac{\sqrt{4x-\kappa_0^2}}{2x} \Str \widetilde{U}\diag(\widehat{S}_{\B\B},\widehat{S}_{\B\B})\widetilde{U}^{-1}\widetilde{\kappa}\right]\frac{d\mu(\widetilde{U})}{\pi^{k^2-k_+^2-k_-^2}}.\nonumber
\end{eqnarray}
The exponential term $\exp[\mathcal{Y}_{\widetilde{U}}]$ is the application of the Rothstein vector field $\mathcal{Y}_{\widetilde{U}}$ which takes care of all Efetov-Wegner boundary terms~\cite{Wegner1983,Efetov1983} that result from the corresponding change of coordinates.

The result~\eref{stable.partition.c} is a superposition of the Poisson partition function~\eref{Poisson.partition.limit} convolved with the stable distribution $\widehat{p}_{\alpha/2}(x)$ and the sine kernel partition function in its supersymmetric form~\cite{Zirnbauer1996,Verbaarschot2004} convolved again with $\widehat{p}_{\alpha/2}(x)$. For the sine kernel, usually the vector field is dropped as it only generates lower point correlations than the $k$-point correlation function that can be derived by taking derivatives in $\kappa$ and then setting $\kappa_{\B\B}=\kappa_{\F\F}$, see~\cite{Efetovbook,Zirnbauer2006,Guhr2011}.

For the computation above, we have assumed that  $\kappa_0$ is of order $\mathcal{O}(1)$ in $N$. We could also choose that $\kappa_0$ is of a larger order since the spectrum has a heavy tail so that eigenvalues can indeed lie very deep in the tail. For instance this is the case for the product of inverse Ginibre matrices where the ratio of the scale of the largest eigenvalue and the bulk is of order ${\rm scale}_{\rm largest\ eigenvalue}/{\rm scale}_{\rm bulk}=N$.

Assuming $\kappa_0\gg 1$ in~\eref{stable.partition.c} with $\widetilde{\kappa}= \mathcal{O}(\kappa_0)$, we can Taylor expand the square root in the first term which makes the exponent $x$ independent so that we get the Poisson partition function~\eref{Poisson.partition.limit},
\begin{equation}
\fl\int_0^{\kappa_0^2/4} dx\widehat{p}_{\alpha/2}(x)\exp\left[-\frac{\kappa_0-\sqrt{\kappa_0^2-4x}}{2x}\Str \widetilde{\kappa}\right]\overset{\kappa_0\gg1}{\approx}\exp\left[-\frac{1}{\kappa_0}\Str\widetilde{\kappa}\right].
\end{equation}
For the second term we rescale $x\to \kappa_0^2 x/2$ then the Jacobian together with the approximation $\widehat{p}_{\alpha/2}(\kappa_0^2 x/2)\propto (\kappa_0^2 x)^{-1-\alpha/2}$  is of the size $\kappa_0^{-\alpha}$ while the integrand is of order one. Thus it is a lower order term and vanishes when $\kappa$ is of an order larger than $\sqrt{N}$.

In summary, for $\kappa\gg\sqrt{N}$, the partition function becomes the one for the Poisson statistic,
\begin{eqnarray}\label{stable.partition.d}
 Z_{\alpha}^{(k,N)}(\kappa)\overset{N\to\infty,\ \kappa\gg\sqrt{N}}{\approx}\exp\left[-\frac{1}{\kappa_0}\Str\widetilde{\kappa}\right]+\mathcal{O}(\kappa_0^{-\alpha}).
\end{eqnarray}
The problem is that the convergence is slower for smaller $\alpha$.

The question is whether the largest eigenvalues are now of larger order than the scale of the macroscopic level density for the ensemble~\eref{GUE.stable}.  Using the averaged position of the largest and smallest eigenvalue in the unfolded variables~\eref{unfolding}, we see that their position (horizontal lines in Fig.~\ref{fig:level-density-stable}) moves to the extreme values at $\pm0.5$. As their change is however very tiny it could be very likely that those positions saturate at a certain value which implies the mixed statistics for the level spacing distribution between the four largest and smallest consecutive eigenvalues seen in Fig.~\ref{fig:Spacing-distributions-stable} will persist when taking the limit $N\to\infty$. We have also simulated the ensemble for $\alpha=0.5$ and $N=5000$ and it seems that this kind of saturation is taking place.
Nonetheless, we can confirm that the similarity to the Poisson statistics is diminished for a smaller $\alpha$ which can be indeed understood by the resulting error term in Eq.~\eref{stable.partition.d}. A more detailed analysis is needed to decide which scenario either~\eref{stable.partition.c} or~\eref{stable.partition.d}, is actually realised.

\section{Conclusions and Two Conjectures}\label{sec:conclusio}

We investigated heavy-tailed unitarily invariant random matrices and their limiting spectral statistics in the tail. In particular we addressed the question whether the statistics are stable when all the remaining statistics are stable. To achieve this we considered two classes of random matrices. One is a product of inverse Ginibre matrices that are known~\cite{BPB1999,APA2009} to yield a freely stable macroscopic level density. Surprisingly in the tail the spectrum is not stable. When $L$ is the number of matrices added in order to check the stability, the eigenvalues in the tail cluster into groups of $L$ eigenvalues. Eigenvalues inside this cluster become statistically independent in the limit of large matrix dimension $N$ while eigenvalues in different clusters are still correlated. Our interpretation is that that the sum of heavy-tailed random matrices behave in the tail like a direct sum of the same types of random matrices. Our analytical computations with the supersymmetry method confirm this. When looking at the particular details of the computation one notices that this easily carries over to a sum of heavy-tailed random matrices that do not necessarily have to be equally distributed nor do they need to have the same stability exponent. We believe that this is even true for real and quaternionic matrices. We now arrive at our first conjecture.

\begin{conjecture}[Tail Statistics of a Sum of Heavy-Tailed Random Matrix Ensembles]\label{conj:sum}
Let $X_1,\ldots,X_L$ be independently (not necessarily identically) distributed random matrices with heavy-tails, and unitary invariance (eigenvalues and eigenvectors are uncorrelated) such that the position of the largest eigenvalues is on a scale larger than that of the eigenvalues in the bulk. Then the statistics of the largest eigenvalues in the tail of the sum $\sum_{j=1}^L X_j$ and of the direct sum $\bigoplus_{j=1}^L X_j$ will be the same up to a scaling.
\end{conjecture}

As we have seen there are also stable statistics in the product of inverse Ginibre matrices. The scale ``${\rm scale}_{\rm critical}$'' of the transition from stable to unstable spectral statistics has been quantified by the following ratio of scales ${\rm scale}_{\rm largest\ eigenvalue}/{\rm scale}_{\rm critical}=N^{1/(2\alpha)}$, where $\alpha$ is the stability exponent. We think that this critical scaling might be universal. Certainly a further investigation is needed for more general classes of random matrix ensembles like the multiplicative P\'olya ensembles~\cite{KK2019,FKK2020}, that also comprise several heavy-tailed ensembles.

It seems to be paramount in Conjecture~\ref{conj:sum} that the largest eigenvalues are on a larger scale than the bulk as we have observed with the second class of ensembles. This class consists of averaged GUE's, where one integrates over the variance with a stable univariate distribution. This construction is very similar to the one in~\cite{BGW2006,BCP2008,AFV2010,AV2008,AMAV2009}, where other averaging distributions have been studied. This led to a heavy-tailed random matrix  that is already for fixed matrix dimension $N$ stable. With the help of this class, we wanted to examine whether the limiting statistics for $L\to\infty$ follows the Poisson statistics as it is the natural choice for a direct sum of random matrices. Our numerical simulations suggest that this is not true. Through our analytical computations we have found out that it has to follow the Poisson statistics if the largest eigenvalue scales much larger than the bulk otherwise one should find a mixture of Poisson statistics and a kind of average of the sine-kernel result. The latter seems to be the case of this average of the GUE with a heavy-tailed standard deviation. Thus, we state our second conjecture.

\begin{conjecture}[Tail Statistics of Stable Random Matrix Ensembles]\label{conj:stable}\

Let $X$ be a heavy-tailed stable random matrix with unitary invariance (eigenvalues and eigenvectors are uncorrelated). If the largest eigenvalues are considerably larger than the bulk of eigenvalues then the local spectral statistics in the tail follows Poisson statistics.
\end{conjecture}

These two conjectures should certainly also carry over in some way to the other symmetry classes for the Hermitian, see~\cite{Zirnbauer1996}, as well as non-Hermitian random matrix ensembles. Surely for complex spectra other mechanisms will enter the game. Nevertheless in two dimensions the added spatial capacity will further facilitate an increase in the decorrelation of eigenvalues.


\section*{Acknowledgments}

MK acknowledges fruitful discussions with Jiyuan Zhang and Holger K\"osters.

\end{document}